\documentclass[11pt]{article}

\makeatletter
\newdimen\xfigwd
\AtBeginDocument{\global\xfigwd=\textwidth}
\makeatother

\usepackage[margin=1in]{geometry}
\usepackage{cite}
\usepackage{amsmath,amssymb,amsfonts}
\usepackage{algorithmic}
\usepackage{graphicx}
\graphicspath{{./figures/}}
\usepackage{textcomp}
\usepackage{xurl}
\usepackage{mathrsfs}
\usepackage{caption}
\usepackage[table,xcdraw]{xcolor}
\usepackage{tabularx}
\usepackage{booktabs}
\usepackage{comment}
\usepackage{float}
\usepackage{bm}
\usepackage[hidelinks]{hyperref}
\usepackage{xcolor}
\usepackage{titlesec}

\definecolor{sectionblue}{RGB}{0,102,204}

\titleformat{\section}
  {\normalfont\Large\bfseries\color{sectionblue}}
  {\thesection}{1em}{}

\titleformat{\subsection}
  {\normalfont\large\bfseries\color{sectionblue}}
  {\thesubsection}{1em}{}

\titleformat{\subsubsection}
  {\normalfont\normalsize\bfseries\color{sectionblue}}
  {\thesubsubsection}{1em}{}

\definecolor{ieeeaccessblue}{RGB}{0,102,204}
\DeclareCaptionLabelFormat{ieeeaccessblue}{\textcolor{ieeeaccessblue}{#1~#2}}
\makeatletter
\AtBeginDocument{%
\DeclareMathVersion{bold}
\SetSymbolFont{operators}{bold}{T1}{times}{b}{n}
\SetSymbolFont{symbols}{bold}{OMS}{cmsy}{b}{n}
\renewcommand\boldmath{\@nomath\boldmath\mathversion{bold}}}
\makeatother

\def\BibTeX{{\rm B\kern-.05em{\sc i\kern-.025em b}\kern-.08em
    T\kern-.1667em\lower.7ex\hbox{E}\kern-.125emX}}

\title{Emulation-based System-on-Chip Security Verification: Challenges and Opportunities}

\title{Emulation-based System-on-Chip Security Verification: Challenges and Opportunities}

\title{Emulation-based System-on-Chip Security Verification: Challenges and Opportunities}

\author{
Tanvir Rahman$^{1}$\thanks{Corresponding author: \texttt{tanvir.rahman@ufl.edu}},
Shuvagata Saha$^{1}$,
Ahmed Y. Alhurubi$^{1}$,\\
Sujan Kumar Saha$^{1}$,
Farimah Farahmandi$^{1}$,
Mark Tehranipoor$^{1}$\\[0.5em]
$^{1}$Department of Electrical and Computer Engineering\\
University of Florida, Gainesville, FL 32611, USA
}

\date{}

\date{}

\date{}

\begin{document}

\maketitle

\begin{abstract}
Increasing system-on-chip (SoC) heterogeneity, deep hardware/software integration, and the proliferation of third-party intellectual property (IP) have brought security validation to the forefront of semiconductor design. While simulation and formal verification remain indispensable, they often struggle to expose vulnerabilities that emerge only under realistic execution conditions, long software-driven interactions, and adversarial stimuli. In this context, hardware emulation is emerging as an increasingly important pre-silicon verification technology because it enables higher-throughput execution of RTL designs under realistic hardware/software workloads while preserving sufficient fidelity for security-oriented analysis.

This paper presents a comprehensive survey and perspective on emulation-based security verification and validation. We organize the landscape of prior work across assertion-based security checking, coverage-driven exploration, adversarial testing, information-flow tracking, fault injection, and side-channel-oriented evaluation. We provide a structured view of emulation-enabled security verification workflows, including instrumentation, stimulus generation, runtime monitoring, and evidence-driven analysis. We also examine practical challenges related to observability, scalability, property specification, and the definition of security-oriented coverage metrics for emulation-based verification. Finally, we discuss emerging directions such as AI-assisted emulation, digital security twins, chiplet-scale security exploration, automated vulnerability assessment, and cloud-scale secure emulation. Overall, this paper positions emulation as a promising foundation for the next generation of pre-silicon hardware security assurance.
\end{abstract}

\vspace{0.5em}

\section{Introduction}
\label{sec:introduction}

System-on-chip (SoC) platforms have evolved into highly integrated systems that combine general-purpose processors, domain-specific accelerators, memory hierarchies, complex interconnects, and a wide range of peripheral interfaces on a single die~\cite{b_SOC,b1,b2}.
This integration is driven by performance, energy efficiency, and time-to-market requirements, but it also increases design complexity and expands the number of security-relevant interactions across hardware, firmware, and system software~\cite{b3,b4}.
Modern SoCs frequently incorporate third-party intellectual property (3PIP), microarchitectural optimizations, and globally sourced design components, all of which complicate security assurance.

Over the past decade, research, surveys, and case studies have documented hardware Trojans, side-channel vulnerabilities, privilege-escalation paths, and other hardware-centric attack vectors affecting both general-purpose and application-specific SoCs~\cite{b5,b6,b7,b8}.
These risks are not merely theoretical.
Publicly disclosed speculative-execution attacks on widely deployed processors, as well as side-channel and fault attacks against deployed cryptographic implementations, have shown that hardware security failures can escape conventional validation and affect real products at scale~\cite{b7,b21,b22,b23,b24}.
Taken together, these results show that security validation must be addressed throughout the design lifecycle, including both pre-silicon and post-silicon stages~\cite{b1,b3}.
Pre-silicon validation is essential because it provides the earliest opportunity to detect and correct architectural and RTL-level weaknesses, whereas post-silicon validation remains necessary because some effects become visible only in silicon or under field conditions.

Traditional verification methodologies provide essential but incomplete coverage for these challenges.
Formal verification and assertion-based checking are effective when the security property is precisely specified and the design scope remains tractable~\cite{b4}.
In practice, these methods are particularly useful for proving localized invariants, such as access-control rules, privilege-transition correctness, and protocol safety properties in selected subsystems~\cite{b9,b4}.
However, formal methods face well-known obstacles at SoC scale, including state-space explosion, the difficulty of specifying complete security intent, and the challenge of modeling realistic firmware and software environments~\cite{b10,b11,b2}.
Simulation-based verification, in turn, provides strong debug visibility and flexibility, but its lower execution speed limits deep state exploration, rare-event discovery, and long-running HW/SW validation campaigns~\cite{b2}.

To address blind spots in these traditional flows, the community has developed several complementary techniques.
Hardware information-flow tracking (IFT), for example, assigns labels to data and propagates them through RTL or gate-level computations to \emph{track} unauthorized flows and enforce confidentiality or integrity policies~\cite{b12,b13,b14}.
Such techniques have been used to study information leakage, privilege escalation, and data-exfiltration scenarios in accelerators and SoC subsystems~\cite{b12,b14}.
Likewise, adversarial testing approaches inspired by fuzzing and penetration testing have been applied to SoC security verification by generating malformed transactions, mutated inputs, and software-driven attack sequences~\cite{b15,b16}.
Side-channel and fault-oriented analyses add another dimension by considering power, electromagnetic (EM), timing, and injected perturbations as sources of exploitable behavior~\cite{b21,b22,b23,b24,b25}.
Together, these approaches broaden security coverage, but they also highlight a common bottleneck: many meaningful security scenarios require long executions, realistic software stacks, and large campaign volume.

Hardware emulation is increasingly important in this setting because it provides a practical way to execute RTL designs at much higher throughput than software simulation while preserving cycle-accurate behavior for the modeled design~\cite{b2,b27,b53}.
Emulation can boot firmware, run operating systems, interact with realistic peripheral models, and sustain workloads long enough to exercise deep system states that are difficult to reach in simulation-only flows.
When paired with security-oriented monitors, targeted tracing, and host-driven test generation, emulation becomes a compelling pre-silicon vehicle for workload-driven and campaign-scale security validation.

Related work has already explored parts of this space, including SoC security validation for specific targets, hardware-assisted adversarial testing, and runtime monitoring techniques that can benefit from high-throughput execution~\cite{b13,b14,b15,b16,b26,b113}.
At the same time, the literature remains fragmented across subsystems, threat models, and validation styles.
What is still missing is a clear and organized view of how emulation fits into SoC security verification as a workflow, what kinds of security checks it enables in practice, and where its main opportunities and limitations lie.

This paper addresses that need by providing a structured perspective on emulation-based security verification and validation.
Our main contributions are as follows:
\begin{itemize}
    \item We clarify the role of hardware emulation within the broader SoC security-verification landscape and explain why it is well suited to long-running, software-intensive, and campaign-scale validation tasks.
    \item We organize existing emulation-enabled security-verification workflows into a concise taxonomy that distinguishes verification intent, workflow structure, and observability choices.
    \item We analyze the main engineering trade-offs that shape practical adoption, including compile and runtime cost, instrumentation overhead, evidence collection, and debug and coverage considerations.
    \item We discuss emerging directions and open research problems that illustrate how emulation can strengthen pre-silicon security assurance for modern SoCs.
\end{itemize}

The remainder of this paper is organized as follows.
Section~II reviews the broader SoC security-verification landscape and the limitations of existing methods.
Section~III introduces hardware emulation and discusses platform types, benefits, and bottlenecks.
Section~IV presents the taxonomy of emulation-based security-verification workflows.
Section~V and Section~VI introduce emulation based security verification and associated challenges respectively. 
Section~VII discusses opportunities and emerging directions, and open research problems.
Finally, Section~VIII concludes the paper.

\section{Overview of SoC Security Verification}
\label{sec:soc_security_overview}

SoC security verification relies on several complementary classes of techniques, including information-flow tracking (IFT), formal verification, simulation-based validation, fuzzing and penetration testing, side-channel and fault analysis, and AI-assisted analysis.
Each class targets different aspects of security assurance and exposes different trade-offs in scalability, precision, and realism.
This section briefly reviews these approaches as background for the later discussion.
Table~\ref{tab:socsec_overview} summarizes the major SoC security-verification approaches discussed in this section, highlighting their main security targets, practical strengths, and recurring bottlenecks at SoC scale.

\subsection{Information-Flow Tracking (IFT)}

Information-flow tracking assigns labels to data (e.g., \emph{secret}, \emph{public}, \emph{trusted}, \emph{untrusted}) and monitors how these labels propagate through the design~\cite{b12,b13,b33}.
IFT is used to detect violations such as leakage of sensitive values to untrusted sinks or unauthorized influence of untrusted inputs on control signals.

IFT can be broadly categorized into:
\begin{itemize}
    \item \textbf{Static IFT (SIFT):} Static approaches analyze RTL, gate-level netlists, or higher-level representations to construct dependence graphs and reason about possible information flows without executing the design~\cite{b12,b33}. They are useful for early vulnerability screening, but they often over-approximate behavior, which can lead to false positives.
    \item \textbf{Dynamic IFT (DIFT):} Dynamic approaches track labels at runtime as the design executes~\cite{b13,b34}. They capture actual execution paths and can expose transient or trigger-dependent behaviors, but they introduce additional runtime and instrumentation overhead.
\end{itemize}

IFT remains attractive because it directly targets confidentiality and integrity policies, but it faces important limitations in large SoCs.
First, label propagation can become expensive as design size and state space grow.
Second, accurately modeling implicit flows and microarchitectural effects is difficult.
Third, policy precision is a recurring challenge: aggressive over-approximation creates false positives, while aggressive pruning risks missing real violations~\cite{b12,b13,b14,b33,b34,b35}.

\subsection{Formal Verification}

Formal verification attempts to prove or refute whether a design satisfies a set of properties for all legal behaviors within a given model~\cite{b9,b10,b11}.
In security verification, formal methods are commonly used to check access-control invariants, privilege-transition correctness, protocol safety, and other properties that can be expressed precisely.

In practice, formal verification is especially effective for localized subsystems and well-defined properties.
Its main strength is exhaustiveness within scope: when a proof succeeds, it rules out an entire class of violations for the modeled design under the stated assumptions.
However, formal methods face well-known barriers in SoCs, including state-space explosion, heavy abstraction effort, and the difficulty of modeling realistic software, peripherals, and environment assumptions~\cite{b9,b10,b11}.
As a result, formal verification is indispensable for targeted security reasoning, but it is rarely sufficient by itself for full-system validation.

\subsection{Fuzzing and Penetration Testing}

Fuzzing applies randomized, mutated, or guided inputs to a design under test to trigger unexpected behaviors and expose bugs~\cite{b15,b16,b26}.
At the hardware level, fuzzing complements IFT and formal methods by exploring large input spaces without requiring explicit models for all possible attack patterns.

Existing hardware fuzzing approaches broadly fall into two categories:
\begin{itemize}
    \item \textbf{Hardware-as-Software (HW-as-SW) fuzzing:} RTL is translated to a cycle-accurate software model, and software fuzzers are applied to this model~\cite{b15,b26}. This allows reuse of mature software fuzzing infrastructure, but performance and abstraction fidelity can be limiting.
    \item \textbf{Hardware-as-Hardware (HW-as-HW) fuzzing:} The design is exercised directly in hardware-oriented environments such as FPGA-based setups or accelerated validation platforms~\cite{b16,b36,b37,b38}. This improves fidelity and throughput, but typically requires more specialized infrastructure.
\end{itemize}

Penetration testing adapts attacker-centric reasoning to pre-silicon validation.
Instead of only asking whether the design is functionally correct, it asks whether an adversary can drive the design into an unsafe or exploitable state under constrained inputs~\cite{b16,b36,b37}.
These methods are effective for exposing adversarial corner cases, but their usefulness depends strongly on the quality of the generated stimuli, the security oracles used for checking, and the practical ability to reproduce and triage interesting failures.

\begin{table}[H]
\centering
\caption{Representative SoC security verification approaches and their main practical bottlenecks.}
\label{tab:socsec_overview}
\renewcommand{\arraystretch}{1.15}
\setlength{\tabcolsep}{5pt}
\small
\begin{tabularx}{\textwidth}{@{}p{3.0cm} p{3.4cm} p{3.4cm} X@{}}
\toprule
\textbf{Approach} &
\textbf{What it targets} &
\textbf{Strengths} &
\textbf{Main practical bottlenecks} \\
\midrule

IFT (static / dynamic) \cite{b12,b13,b14,b33} &
Confidentiality/integrity flow policies across RTL and HW/SW boundaries. &
Finds unintended propagation; aligns with security-policy checking; actionable alerts. &
\begin{itemize}\itemsep0.1em
  \item Taint/state explosion on large SoCs
  \item Instrumentation overhead and runtime cost
  \item Precision vs.\ false-positive trade-offs
\end{itemize}
\\

Formal verification \cite{b9,b10,b11} &
Proof/violation of security properties (invariants, ordering, access control, protocol safety). &
Exhaustive for bounded scopes; strong assurance for targeted properties. &
\begin{itemize}\itemsep0.1em
  \item State explosion; abstraction effort
  \item Hard to model full-system software behaviors
  \item Limited practicality for long system traces
\end{itemize}
\\

RTL simulation validation &
Workload-driven testing for security checks, regressions, and bug reproduction. &
Mature debug; deterministic replay; low entry cost. &
\begin{itemize}\itemsep0.1em
  \item Slow for firmware/OS boot and long runs
  \item Limited iterations for large campaign testing
  \item High cost for broad configuration sweeps
\end{itemize}
\\

Fuzzing-based testing \cite{b15,b17,b18} &
Coverage-guided or randomized stimulus to trigger security-relevant behaviors. &
Finds deep corner cases; automation-friendly; effective for stress testing. &
\begin{itemize}\itemsep0.1em
  \item Input throughput can be limited by execution speed
  \item Fidelity vs.\ observability tension
  \item Expensive multi-week campaigns in slow environments
\end{itemize}
\\

Attack-oriented validation \cite{b16} &
Adversarial scenarios (protocol abuse, malformed transactions, misuse cases). &
Connects design bugs to exploitability; validates mitigations end-to-end. &
\begin{itemize}\itemsep0.1em
  \item Many scenarios require long executions
  \item Hard to scale across versions/configurations
  \item Reproduction + triage cost dominates
\end{itemize}
\\

Side-channel / fault evaluation \cite{b21,b22,b24,b25} &
Leakage/fault sensitivity via measurement or models (power/EM/timing, fault campaigns). &
Captures non-functional vulnerabilities; validates countermeasures. &
\begin{itemize}\itemsep0.1em
  \item High data volume; repeated replays
  \item Measurement noise / environment variability
  \item Difficult pre-silicon fidelity for some effects
\end{itemize}
\\

AI/LLM-assisted analysis \cite{b114,b115,b116,b117} &
Triage, artifact understanding, and security reasoning over large SoC contexts. &
Reduces manual effort; supports prioritization and analysis at scale. &
\begin{itemize}\itemsep0.1em
  \item Depends on high-quality traces/data
  \item Sensitive to weak ground truth
  \item Vulnerable to misleading or incomplete evidence
\end{itemize}
\\

\bottomrule
\end{tabularx}
\end{table}

\subsection{Side-Channel and Fault Analysis}

Side-channel analysis and fault-injection assessment evaluate implementations under physical or timing perturbations rather than purely functional tests~\cite{b21,b22,b24,b25}.
Even when a cryptographic algorithm is mathematically sound, implementation details can leak information or create exploitable failure modes.

\textbf{Side-channel analysis} focuses on observables such as power, electromagnetic (EM) emissions, and timing:
\begin{itemize}
    \item Power and EM measurements can be used in attacks such as Differential Power Analysis (DPA) to correlate leakage with secret keys~\cite{b21,b24}.
    \item Timing side channels arise when execution latency depends on secret data or privileged state~\cite{b7,b22}.
\end{itemize}

\textbf{Fault injection} introduces transient or permanent disturbances using techniques such as voltage/clock glitches or laser/EM pulses~\cite{b23,b24,b25}.
By perturbing hardware during cryptographic operations or security-critical transitions, attackers may bypass checks, extract secrets, or corrupt control flow.
Pre-silicon security verification must therefore reason about fault resilience and the effectiveness of protection mechanisms such as parity, redundancy, and fault sensors.

Although many side-channel and fault studies are post-silicon, pre-silicon analyses increasingly rely on models and approximations to identify high-risk structures and evaluate candidate countermeasures~\cite{b21,b22,b23,b24,b25,b44,b120,b121}.

\subsection{AI-Based Security Verification}

AI and machine learning (ML) techniques are being explored to automate anomaly analysis and security reasoning in hardware designs~\cite{b42,b49,b50,b118,b119}.
Representative uses include:
\begin{itemize}
    \item \textbf{Trojan and anomaly detection:} learning-based classifiers trained on structural or trace-based features,
    \item \textbf{Side-channel analysis:} deep models that process noisy power or EM traces,
    \item \textbf{Behavioral modeling:} unsupervised and semi-supervised methods that learn expected hardware behavior and flag deviations.
\end{itemize}

These methods can reduce manual effort and help prioritize suspicious behaviors, but their effectiveness depends heavily on the quality of data, labels, and evaluation methodology~\cite{b42,b49,b50,b118,b119}.

\subsection{Summary and Remaining Gaps}

As summarized in Table~\ref{tab:socsec_overview}, the techniques above collectively form a broad toolbox for SoC security verification, but each has practical limitations.
Formal and IFT-based approaches can become difficult to scale as system complexity grows.
Simulation-based methods provide strong visibility but limited throughput.
Fuzzing and attack-oriented validation can expose deep bugs, but campaign cost and triage effort remain significant.
Side-channel and fault analyses capture important non-functional vulnerabilities, yet many effects are difficult to model faithfully before silicon.
These gaps motivate the search for verification approaches that can combine realistic execution context, long-running behavior, and practical campaign scale.


\section{Hardware Emulation for SoC Security Verification}
\label{sec:emulation_basics}

Hardware emulation executes an RTL design on hardware-assisted platforms to achieve substantially higher throughput than software RTL simulation while retaining cycle-accurate behavior for the modeled design.
In practice, both FPGA prototypes and commercial emulation systems typically operate at \emph{MHz-scale} effective speeds for large SoCs, although achievable frequency depends on design size, resource partitioning, instrumentation, and the selected debug or trace configuration~\cite{b27,b53}.
This section introduces emulation fundamentals, major platform types, and the trade-offs that matter for security verification.

\begin{figure*}[t]
    \centering
    \includegraphics[width=0.70\linewidth]{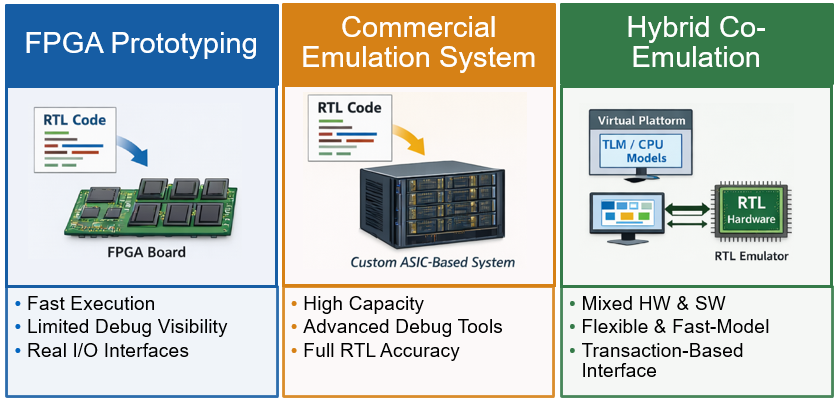}
    \caption{Types of emulation platform: FPGA prototyping, ASIC-class commercial emulation, and hybrid co-emulation; each offers trade-offs in speed, capacity, visibility, and deployment.}
    \label{fig:emu_types}
\end{figure*}

\subsection{Definition and Significance}

In an emulation flow, the design under test (DUT) is compiled (synthesized and partitioned) to a hardware execution fabric.
For FPGA prototyping, the fabric is one or more commodity FPGAs; for commercial emulation, the fabric is a vendor-provided hardware system integrated with dedicated compilation, transactor, and debug infrastructure~\cite{b27,b53,b85,b86,b87}.
After configuration, the DUT runs under host-provided stimulus and can execute long system workloads that are impractical in simulation-only settings.

For security verification, emulation is significant because it enables:
\begin{itemize}
    \item \emph{long-duration} execution (boot flows, OS services, driver activity, extended regressions) where security-relevant behaviors may manifest only after deep traces;
    \item \emph{cross-layer} HW/SW validation by running realistic firmware and software stacks on the pre-silicon model, including interrupts, DMA, and peripheral concurrency;
    \item \emph{campaign-scale} adversarial testing (e.g., robustness campaigns and fault-injection sweeps) where practical effectiveness depends on high iteration counts and repeatability~\cite{b27,b53}.
\end{itemize}
Emulation therefore complements formal verification by scaling dynamic validation to realistic system context, while serving as a higher-throughput alternative to simulation for long-running, software-intensive scenarios.

\subsection{Types of Emulation Platforms}
\label{subsec:emu_platforms}

Emulation-based validation is commonly realized in three deployment categories: FPGA-based prototyping, commercial (ASIC-class) emulation systems, and hybrid co-emulation that couples a virtual platform with an emulator through transaction-level co-modeling~\cite{b27,b53,b110,b111,b112}.
All three execute security-relevant workloads at far higher throughput than RTL simulation, but they differ in capacity, observability, and how much of the system is modeled at RTL fidelity.

\subsubsection{FPGA-Based Prototyping}

FPGA prototyping maps the SoC RTL onto one or more commodity FPGAs.
It is widely used for early HW/SW integration because it provides hardware-speed execution for large RTL blocks, supports real I/O connectivity, and enables software bring-up on a cycle-accurate model~\cite{b27,b53}.
For security evaluation, FPGA prototypes can sustain long-running workloads and realistic peripheral interaction, which is valuable for studying protocol misuse, Trojan triggers, and leakage-sensitive execution paths under near-real conditions~\cite{b21,b24}.

The main practical trade-off is observability: internal probing and trace capacity are constrained, and large designs may require careful partitioning and clocking choices.
As a result, prototyping often emphasizes system-level execution and functional realism, while deep debug is performed selectively or deferred to higher-visibility environments.

\subsubsection{Commercial (ASIC-Class) Emulation Systems}

Commercial emulation systems are enterprise platforms designed for full-chip RTL capacity with integrated compilation, transactors, assertion support, and scalable debug.
In practice, these systems are provided as vendor-optimized emulation platforms with dedicated compilation, visibility, and workflow infrastructure~\cite{b85,b86,b87,cad11,hybrid}.
Their primary advantage is scale: they support very large SoCs together with controlled observability mechanisms such as triggers, selective trace, and save/restore, making them suitable for long regressions and campaign-style validation~\cite{b85,b86,b87,cadence2}.

A common point of confusion is the distinction between \emph{compile/mapping time} and \emph{execution time}.
Mapping can take non-trivial time for large SoCs, especially with heavy instrumentation; however, once an image is built, it can be reused across many hours or days of high-throughput execution, regressions, and large security campaigns.
This amortization is a key reason emulation is used when validation requires long workloads or many iterations~\cite{b85,b86,b87,cad11}.

In practice, these platforms are often coupled with host-driven transactors, accelerated verification IP, and simulation--emulation integration flows, allowing stimulus generation and parts of the testbench environment to remain on the host while the RTL design executes on the emulation hardware~\cite{report1,cad11}.

\begin{figure*}[t]
\centering
\includegraphics[width=0.6\textwidth]{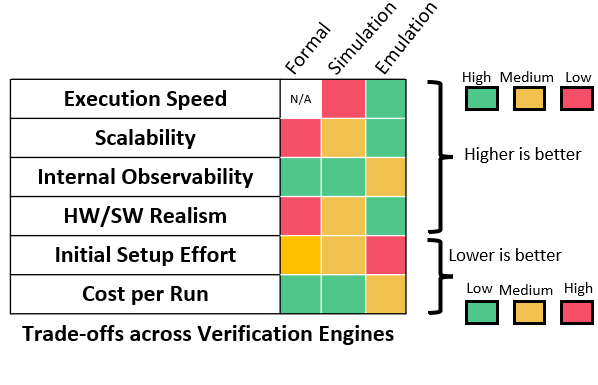}
\caption{Pre-silicon verification spectrum: qualitative comparison of practical trade-offs across verification engines.}
\label{fig:quant_comp}
\end{figure*}

\subsubsection{Hybrid Co-Emulation (Virtual Platform + Emulator)}

Hybrid co-emulation partitions the system across two execution domains: (i) a virtual platform on the host (e.g., high-level SystemC/TLM models or fast CPU/ISA models) and (ii) an emulator executing selected RTL blocks at cycle accuracy.
The two domains are connected through transaction-level co-modeling channels so that software workloads can execute quickly while security-critical hardware behavior remains at RTL fidelity~\cite{b110,b111,b112}.

Hybrid co-emulation is most useful when (a) full-SoC RTL emulation is unnecessarily expensive for early software bring-up, or (b) only a subset of blocks require repeated, high-fidelity security evaluation.
Typical motivations include:
\begin{itemize}
    \item \textbf{Capacity and instrumentation overhead:} reduce the RTL footprint placed on the emulator by keeping well-modeled components virtual while retaining security-critical components in RTL~\cite{b111,b112}.
    \item \textbf{Model maturity asymmetry:} leverage mature virtual models (often CPUs and peripherals) while focusing RTL fidelity on blocks where security behavior is subtle or insufficiently modeled (e.g., interconnect protections, security monitors, DMA/security boundaries)~\cite{b110,b111}.
    \item \textbf{Throughput vs.\ fidelity balance:} accelerate end-to-end workloads while preserving cycle-accurate evaluation where timing, ordering, and microarchitectural behavior matter for security checks~\cite{b110,b112}.
\end{itemize}

For security verification, hybrid co-emulation enables repeated testing of security-relevant RTL components under realistic software execution without always requiring full-chip RTL to be mapped at maximum visibility.
In later stages, the same workflow can be tightened by moving additional blocks from virtual to RTL as needed, trading throughput for fidelity in a controlled manner~\cite{b110,b111,b112}.
Figure~\ref{fig:emu_types} illustrates the main classes of emulation platform.

\subsection{Role of Emulation in the Pre-Silicon Verification Spectrum}

Emulation occupies an important middle ground in pre-silicon security verification, combining RTL-level fidelity with execution speeds high enough to support realistic firmware- and software-driven validation.
In contrast to engines primarily optimized for exhaustive reasoning or fine-grained debug, emulation is especially valuable when security assessment depends on long execution horizons, rich HW/SW interaction, and repeated campaign-scale experimentation.

From a security-validation perspective, three practical characteristics make emulation particularly relevant:
\begin{itemize}
    \item \textbf{Workload depth and realism:} Emulation can execute realistic firmware and OS stacks, I/O activity, and long stress workloads, which are often necessary to expose vulnerabilities that depend on sequencing, timing, privilege transitions, or concurrency.
    \item \textbf{Campaign scale:} Robustness assessment, perturbation sweeps, and security stress testing may require very large numbers of runs; emulation makes such campaign-scale evaluation practical while preserving repeatability.
    \item \textbf{System-context retention:} By keeping the hardware and software context intact during execution, emulation enables security evaluation under operating conditions that more closely resemble deployment than isolated block-level analysis.
\end{itemize}

These strengths make emulation well suited to system-level security validation, particularly when the objective is to evaluate behavior under realistic execution context rather than only local functional correctness.
At the same time, emulation remains subject to practical constraints, including compile and mapping overhead, limited internal visibility, and the cost of moving and storing debug evidence, which makes selective instrumentation and disciplined data collection essential~\cite{b85,b86,b87,cadence2,hybrid}.

\begin{figure*}[t]
    \centering
    \includegraphics[width=0.99\linewidth]{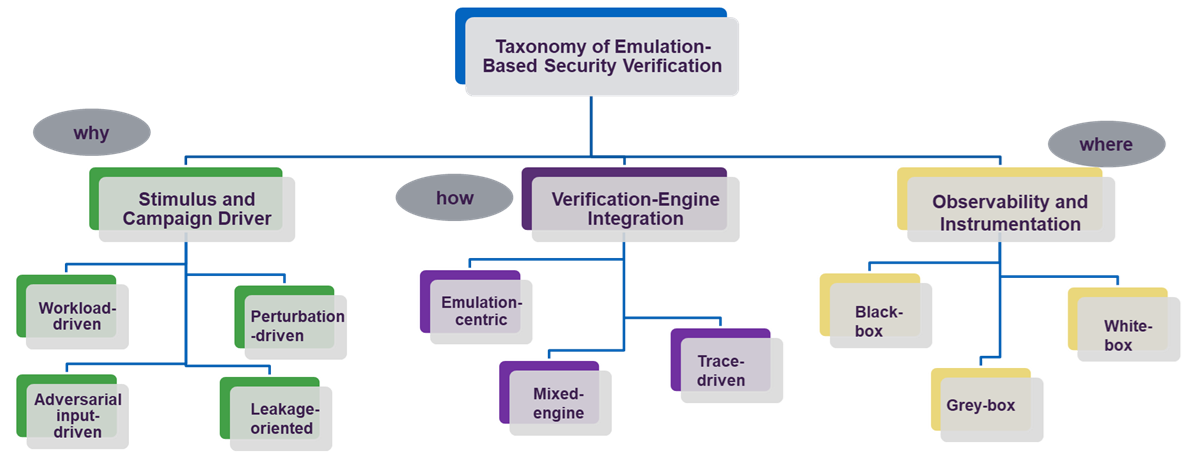}
    \caption{Workflow taxonomy for emulation-based security verification (ESV). The taxonomy classifies ESV workflows by (i) stimulus/campaign driver, (ii) verification-engine integration, and (iii) observability/instrumentation. This taxonomy separates workflow choices from emulation \emph{platform} types (Section~III.B).}
    \label{fig:taxonomy_tree}
\end{figure*}

Figure~\ref{fig:quant_comp} qualitatively positions emulation within the broader pre-silicon verification spectrum.
Rather than serving as a benchmark, it summarizes commonly observed workflow-level trade-offs across execution throughput, scalability, observability, HW/SW realism, setup effort, and per-run cost, based on prior discussions of logic emulation, vendor platform capabilities, and hybrid co-emulation practice~\cite{b53,b85,b86,b87,b110,b111,b112,cad11,hybrid}.
In this spectrum, emulation is most beneficial when security validation requires long-running, software-intensive, and campaign-scale execution under realistic system conditions.


\section{Taxonomy of Emulation-Based Security Verification}
\label{sec:taxonomy}

Emulation-based security verification (ESV) spans a family of workflows that execute RTL at high throughput while checking security-relevant behavior under realistic HW/SW activity.
Section~III characterizes \emph{emulation platforms} (e.g., FPGA prototyping, commercial emulation systems, and hybrid co-emulation) and their trade-offs in capacity, speed, and debug visibility (Figure~\ref{fig:emu_types}).
This section instead classifies the \emph{verification workflow} executed on top of an emulation platform: how security-relevant execution is generated, how checking is performed, and what evidence is collected for triage and debug.
Keeping these lenses separate helps avoid confusion between ``platform types'' and ``verification types.''

We organize ESV methods along three workflow dimensions:
\begin{enumerate}
    \item \textbf{Stimulus and campaign driver:} how execution is generated to expose security-relevant behavior.
    \item \textbf{Verification-engine integration:} how emulation is incorporated into a broader verification stack.
    \item \textbf{Observability and instrumentation:} what internal visibility is used and where checking occurs.
\end{enumerate}
An ESV pipeline can be described by selecting one option from each dimension (e.g., \emph{perturbation-driven} stimulus + \emph{emulation-centric execution} + \emph{gray-box monitoring})~\cite{b23,b62,b63,b85}.

\noindent\textbf{How to read the taxonomy.}
The three dimensions correspond to practical design questions that recur in ESV deployments.
\emph{Why} a campaign is run and what behavior it aims to expose is captured by the \textbf{stimulus/campaign driver}.
\emph{How} execution and checking are distributed across verification engines is captured by \textbf{verification-engine integration}.
\emph{Where} checking occurs and what visibility is relied upon is captured by \textbf{observability/instrumentation}.
Finally, \emph{when} each engine is emphasized (early bring-up, campaign-scale execution, or post-trigger localization) is typically determined by the chosen integration mode and evidence strategy.

\subsection{Classification by Stimulus and Campaign Driver}
In ESV, the campaign driver determines what behaviors are explored and at what scale (workload depth, iteration count, and evidence volume).
We group stimulus into four practical categories:

\subsubsection{Workload-driven system execution}
Execute representative firmware, OS, and driver workloads (boot, interrupts, DMA, peripheral activity) and check security properties during realistic operation~\cite{b27,b53,b67}.

\subsubsection{Adversarial input-driven campaigns}
Generate malformed or adversarial transactions or sequences at high volume (fuzzing- or penetration-style testing) to surface unexpected state transitions and unsafe handling paths~\cite{b16,b42,b43,b44,b86}.

\subsubsection{Perturbation-driven campaigns}
Inject controlled disruptions (e.g., RTL fault models, activation-window sweeps, or scheduling/timing stress) and observe whether security expectations are violated~\cite{b23,b62,b63}.

\subsubsection{Leakage-oriented stress and proxies}
Run leakage-sensitive workloads and monitor cycle-level indicators or software-visible symptoms that correlate with timing or state leakage, serving as pre-silicon proxies for side-channel risk~\cite{b21,b22,b24,b25}.

\subsection{Classification by Verification-Engine Integration}
ESV workflows differ in how emulation is combined with other engines for execution, checking, and debug.
To avoid conflating workflow choices with platform types (Section~III), the categories below describe \emph{workflow composition}, not hardware infrastructure.

\subsubsection{Emulation-centric execution}
In this mode, the emulator is the primary execution engine for workloads and campaigns, and checking is performed by synthesizable on-platform monitors (assertions, checkers, and triggers) and/or host-side monitors driven by exported transactions or events~\cite{b27,b53,report1}.

\subsubsection{Mixed-engine workflows}
In this mode, responsibilities are partitioned across engines: short, high-visibility scenarios may be validated in simulation; system-scale execution and campaign volume are handled in emulation; and optional formal analysis is used to validate localized invariants or monitors.
Each engine is used where it is most effective: visibility for short scenarios, throughput for long scenarios, and exhaustiveness for bounded proofs~\cite{b67,b44,b20}.

\subsubsection{Trace-driven localization}
In this mode, emulation is used to execute at scale and collect compact evidence (events, triggers, or short trace windows), while detailed localization is performed offline or in a higher-visibility environment using the collected evidence.
This mode is common when full waveform capture is infeasible during long runs and when diagnosis requires narrowing the failure window before deeper analysis~\cite{b67,hybrid,syn}.

\subsection{Classification by Observability and Instrumentation Strategy}
A key workflow decision is \emph{what} is observed and \emph{where} checking occurs, which strongly affects overhead and scalability~\cite{b27,b37}.
We summarize observability using the widely adopted black-/gray-/white-box lens~\cite{b85}.

\subsubsection{Black-box monitoring}
Black-box ESV checks security-relevant behavior only at externally visible interfaces (architectural outputs, memory-mapped I/O, or bus transactions).
This minimizes instrumentation cost but can miss internal policy violations that do not immediately surface at interfaces~\cite{b85,b37}.

\subsubsection{Gray-box instrumentation}
Gray-box ESV instruments selected internal signals and events (e.g., privilege bits, security state machines, tags, or specific interconnect events) and exports compact traces or counters~\cite{b37,rad1}.
This provides a practical balance between scalability and diagnostic power and is common in large SoC campaigns where full visibility is infeasible~\cite{b27,b53}.

\subsubsection{White-box, property-centric monitoring}
White-box ESV embeds assertions and property monitors throughout the RTL so that violations are detected close to their source~\cite{b53}.
This improves diagnostic precision but consumes emulator resources and can increase compile and mapping overhead, making automation of monitor insertion and trace selection important for scalability~\cite{b27,b37,rad1}.

\subsection{Summary}
ESV is best viewed as a workflow design space rather than a single technique, as summarized in Figure~\ref{fig:taxonomy_tree}.
Separating \emph{platform type} (Section~III.B) from the workflow dimensions above enables more consistent comparison across prior work: the campaign driver explains how behaviors are explored, the integration style explains how emulation fits into the broader verification stack, and the observability choice explains the cost/diagnostic trade-off for evidence collection and debug.


\section{Emulation-Based Security Verification}
\label{sec:esv}

\subsection{General Flow of ESV}

\begin{figure*}[t]
    \centering
    \includegraphics[width=0.85\linewidth]{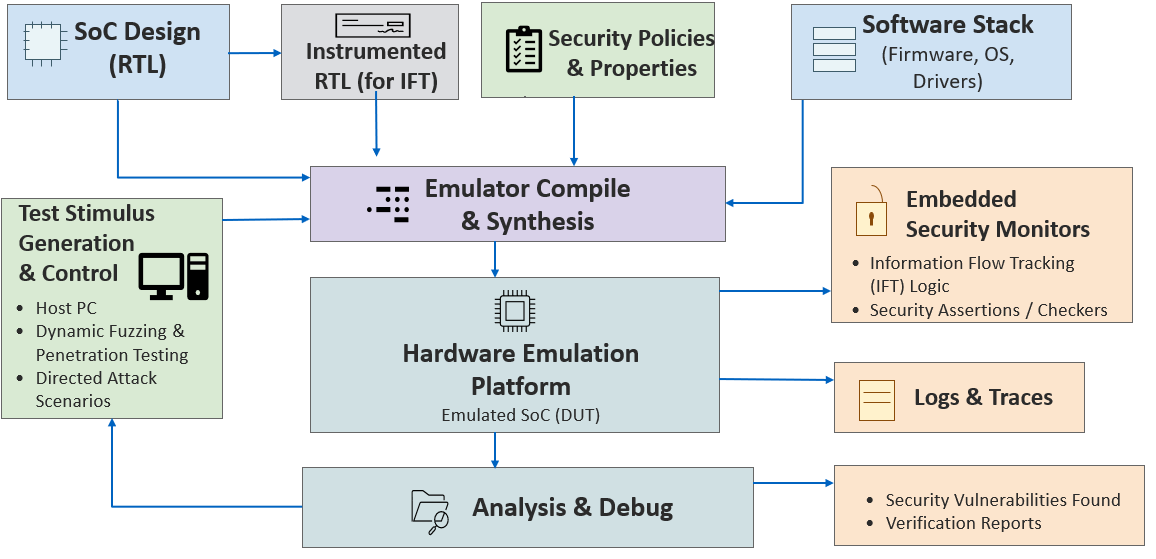}
    \caption{End-to-end emulation-based security verification flow: compile RTL into an emulation platform, run realistic HW/SW workloads, instrument and observe internal signals, and detect security violations via monitors, assertions, and trace analytics.}
    \label{fig:emu_security_flow}
\end{figure*}

A typical ESV campaign follows the high-level flow illustrated in Figure~\ref{fig:emu_security_flow}:

\begin{enumerate}
    \item \textbf{Security requirements and threat modeling:} Security engineers identify critical assets (e.g., keys, credentials, privileged registers), attacker capabilities, and attack surfaces (e.g., external interfaces, DMA engines, debug infrastructure, and third-party IP). ...This step produces a set of security goals and threat scenarios that the ESV campaign must exercise~\cite{b115,b117}.

    \item \textbf{Model preparation and emulation bring-up:} The RTL is synthesized, partitioned, and mapped to the emulator or FPGA platform. At this stage, additional instrumentation may be inserted---including on-chip logic analyzers, counters, status registers, and security monitors---to improve observability and controllability during security campaigns~\cite{b62,b63}.

    \item \textbf{Security instrumentation and monitors:} Hardware information-flow tracking (IFT) logic, security assertions, and rule-based checkers are integrated into the emulation model or tightly coupled monitor blocks. These mechanisms enforce policies such as ``secret data must never appear at untrusted outputs'' or ``user mode must never write to privileged control registers,'' and are configured to raise flags when violations occur~\cite{b37,b38,b115,b116}.

    \item \textbf{Test and attack campaign generation:} Security-oriented workloads are created by combining directed tests, constrained-random sequences, coverage-guided fuzzing, penetration-style scripts, and increasingly, reinforcement learning-guided exploration~\cite{b42,b43,b44,b50,b51}. Cost functions and security oracles translate campaign goals into actionable feedback for these generators.

    \item \textbf{Emulation execution and data collection:} The generated campaigns are executed on the emulation platform. During execution, monitors record assertion failures, IFT violations, protocol errors, fault effects, and other security-relevant events. Logs and traces are collected through on-platform buffers, debug interfaces, or host-side APIs~\cite{b42,b43,b15,b61}.

    \item \textbf{Analysis, diagnosis, and design fixes:} Detected anomalies are classified as real vulnerabilities, benign corner cases, or modeling artifacts. Confirmed issues are traced back to specific modules, interfaces, or microarchitectural behaviors; fixes are then implemented and validated by re-running the relevant ESV campaigns~\cite{b50,b51}.
\end{enumerate}

This flow highlights how emulation serves as the execution backbone that connects threat modeling, embedded security monitors, adversarial stimulus generation, and iterative diagnosis.

\subsection{Core Building Blocks of ESV}

Within this flow, ESV deployments typically integrate four core building blocks: (i) embedded monitors, (ii) dynamic and adversarial stimulus engines, (iii) fault and resilience evaluation, and (iv) cross-layer analysis.

\subsubsection{Embedded Monitors and Property Checkers}

Embedded monitors encode security policies directly in hardware, or in hardware-adjacent logic, and evaluate them at runtime. Examples include:

\begin{itemize}
    \item \textbf{IFT-based monitors}, which propagate taint labels and flag illegal flows from secret sources to untrusted sinks, or from untrusted inputs to privileged state~\cite{b37,b38};
    \item \textbf{Assertion-based security checkers}, which enforce invariants on privilege modes, access-control rules, and control-flow transitions;
    \item \textbf{Rule-based checkers}, which instantiate libraries of common security properties (e.g., CWE-style rules) as synthesizable monitors.
\end{itemize}

Running these monitors on emulation platforms allows them to be stressed under realistic workloads and adversarial traffic, turning them into continuous security sentinels rather than static checklist items.

\subsubsection{Dynamic Fuzzing and Penetration Testing on Emulators}

Fuzzing and penetration testing use emulation as a high-throughput execution engine to explore the design's state space under adversarial stimuli. Hardware-centric fuzzers such as SoCFuzzer, TaintFuzzer, and FormalFuzzer generate bus transactions, protocol sequences, or software inputs while using structural or security-oriented metrics as guidance~\cite{b42,b43,b44}. Processor-level frameworks such as ProcessorFuzz leverage architectural and microarchitectural feedback (e.g., coverage of control and status registers) to expose deep bugs in CPU cores~\cite{b61}.

Penetration-style frameworks such as Sharpen and Re-Pen cast SoC security validation as an optimization or reinforcement-learning problem: the environment is the emulated SoC, the action space is defined by possible attacker operations, and the reward encodes the success of a violation, such as unauthorized access or bypass of a protection mechanism~\cite{b50,b51}. Emulation provides the throughput and HW/SW realism needed for this style of dynamic, attacker-centric validation.

\subsubsection{Fault Injection and Resilience Evaluation}

Emulation is also a natural platform for pre-silicon fault-injection studies. By selectively flipping bits in registers, corrupting memory locations, or perturbing control and status signals, fault frameworks can evaluate whether the design's error-detection and recovery mechanisms are sufficient and whether injected faults can be escalated into security exploits~\cite{b62,b63}.

Security-focused frameworks such as SoFI use property-driven analyses to identify fault-sensitive nodes and prioritize injection locations and timings~\cite{b63}. Emulation's speed enables large fault campaigns and systematic exploration of the fault space that would be prohibitively slow in simulation.

\subsubsection{Cross-Layer and Side-Channel-Oriented Analyses}

Finally, ESV can be extended to support cross-layer and side-channel-oriented analyses. Emulation traces, including switching activity, performance counters, and timing distributions, can be exported and post-processed as proxies for power or timing behavior, allowing early evaluation of side-channel countermeasures and microarchitectural isolation schemes~\cite{b15,b37}. Similarly, emulation can be used to validate whether mitigations for speculative execution, cache side channels, or DMA-based attacks behave as intended under complex software and concurrent workloads~~\cite{b7,b27,b32}.

\subsection{Positioning ESV in the Overall Verification Landscape}

Viewed within the broader SoC verification landscape, ESV occupies a middle ground between purely formal or simulation-based methods and post-silicon evaluation. It offers:

\begin{itemize}
    \item the \emph{scalability and realism} needed to execute full software stacks and long adversarial campaigns;
    \item the \emph{instrumentation flexibility} to integrate IFT logic, assertions, and fault hooks that would be too intrusive or costly in silicon;
    \item the \emph{feedback mechanisms} needed to support modern fuzzing, penetration testing, and RL-guided exploration engines~\cite{b42,b43,b44,b50,b51,b61}.
\end{itemize}

The next sections build on this view of ESV to organize emulation-based security workflows more systematically and to discuss the key opportunities and challenges involved in bringing ESV closer to mainstream industrial practice.

\section{Challenges in Emulation-Based Hardware Security Verification}
\label{sec:challenges}

As shown in Figure~\ref{fig:overhead_mix}, deploying emulation-based security verification requires overcoming significant resource and observability constraints. Key challenges include managing the logic and routing overhead of security monitors, adapting attack harnesses to emulator timing and capacity limits, and strategically selecting internal signals for analysis within finite trace-buffer budgets. Addressing these factors is essential for making emulation-based security verification scalable, automatable, and practically usable in verification sign-off flows.

\begin{figure*}[t]
    \centering
    \includegraphics[width=0.99\linewidth]{figure_challengepng.png}
    \caption{Critical technical dimensions and overheads in emulation-based security verification. This map outlines architectural and methodological challenges---including scaling, instrumentation density, and trace-buffer constraints---that must be addressed to achieve scalable and automated security sign-off.}
    \label{fig:overhead_mix}
\end{figure*}

Despite its unique advantages in speed, capacity, and realism, emulation-based security verification is far from a push-button solution. Practical deployments must contend with substantial resource constraints, methodological friction, and gaps in how security results are observed and quantified. Building on prior work on logic emulation and emulation-based verification~\cite{b27,b53}, as well as recent security-centric frameworks~\cite{b15,b49,b62,b63}, we group the main challenges into three broad categories: (i) computational overheads and cost constraints, (ii) integration with existing verification workflows, and (iii) debuggability, observability, and coverage.

\subsection{Computational Overheads and Cost Constraints}

Modern emulators and large FPGA-based prototyping platforms are capital-intensive infrastructure. Acquisition and maintenance costs effectively restrict full-featured emulation to a limited number of large design houses and semiconductor vendors~\cite{b27,b53,b85,b86,b87}. Even within such organizations, emulator slots are often shared across multiple projects and teams, so emulation must be scheduled and prioritized carefully~\cite{b27,b53,b85,b86,b87}.

From a technical standpoint, mapping a security-instrumented SoC onto an emulator is non-trivial. Studies of large FPGA-based logic emulation systems report that synthesis, partitioning, and compilation for complex designs can take many hours or even days~\cite{b27}. Additional security instrumentation---such as fault-injection fabrics, information-flow monitors, and fuzzing harnesses---further increases logic utilization and routing pressure, exacerbating compile and bring-up time. Similar challenges are reported in industrial emulation flows, where verification teams must budget compilation time and emulator usage alongside functional regressions and software co-validation~\cite{b27,b53,b85,b86,b87}.

These overheads directly affect how aggressively emulation can be used for security. Security verification workloads often require very long execution windows (e.g., billions of cycles for fuzzing or systematic fault campaigns) and may need to be repeated under multiple threat models or fault configurations~\cite{b15,b49,b62,b63}. In practice, this creates a tension between functional and security objectives: emulator capacity is often reserved for tape-out--critical functional regressions, and security campaigns may be restricted to a subset of high-risk blocks or deferred to later project stages. FPGA-based prototyping can mitigate acquisition costs, but multi-FPGA partitioning, clock-domain management, and limited debug visibility introduce additional engineering overheads and can be particularly challenging for complex SoC security scenarios~\cite{b27,b53,b87}.

\subsection{Integration with Existing Verification Workflows}

A second major challenge is methodological integration. Industrial verification flows are typically optimized for functional coverage closure; emulation is already used to accelerate software bring-up and system validation, but security-specific workloads are rarely treated as first-class citizens in these environments~\cite{b27,b53,b85,b86,b87}. Threat models, security properties, attack oracles, and coverage objectives must be woven into existing regression and sign-off processes, which were not originally designed with adversarial behaviors in mind.

Bridging simulation and emulation environments is particularly complex. Unified verification flows have been proposed in which UVM testbenches and verification IP are accelerated by mapping parts of the environment onto the emulator while keeping other components in software~\cite{b53,report1}. However, security monitors and attack harnesses must often be re-engineered to meet emulator timing and capacity constraints, leading to divergent environments between simulation and emulation. This divergence can introduce subtle inconsistencies in observed behavior, complicating debug and making it more difficult to reproduce security issues across platforms.

Automation is another pain point. In simulation, regression-management tools routinely schedule thousands of tests, perform coverage analysis, and track closure progress. Extending these tools to reason about security goals and to schedule long-running emulation campaigns is non-trivial; emulator jobs must be co-optimized with functional runs, and security tests may require coordinated software stacks (e.g., secure boot, firmware, and OS services), which increases orchestration complexity. Advanced vendor methodologies for emulation---including accelerated UVM and unified simulation--emulation environments~\cite{report1,hybrid}---provide a starting point, but security-specific workflows (e.g., hardware fuzzing, structured fault campaigns, and property-driven attack exploration) remain at an early stage of integration.

\subsection{Debuggability, Observability, and Coverage}

Although emulation offers significantly better internal visibility than post-silicon bring-up, debuggability and observability remain key bottlenecks when the goal is to root-cause subtle security violations. Capturing every internal signal at full emulation speed is infeasible; waveform and trace buffers are finite, and storing all activity over billions of cycles is prohibitive. Consequently, verification engineers must select a subset of signals or trigger conditions \emph{a priori}. If a vulnerability arises from an unmonitored interaction, the design may need to be recompiled with additional probes, incurring the same long compilation overhead discussed earlier~\cite{b27,cadence2,hybrid}.

Security-centric bugs amplify this challenge. Many hardware security violations are functionally silent: a confidentiality breach may not corrupt outputs, and an integrity violation might only briefly violate a security property before being masked. Detecting such issues often requires correlating multiple sources of evidence, such as assertion violations, fuzzing feedback, information-flow alerts, firmware logs, and side-channel proxies. Traditional debug infrastructures for emulation---centered on transaction logs, limited waveform capture, and static trigger conditions---were not designed for this kind of multi-modal forensic analysis~\cite{cadence2,hybrid}. Recent industrial advances in replay-, save/restore-, and app-assisted debug for emulation and prototyping begin to address these limitations~\cite{hybrid,syn,cadence2}, but integrating such capabilities into security verification flows in a systematic way remains an open problem.

Coverage adds a further dimension of difficulty. Emulation platforms are well suited to running long tests and exploring large state spaces, but there is still no widely accepted notion of ``security coverage'' for such campaigns. Functional metrics such as line, toggle, or FSM coverage correlate only loosely with an attacker's search space. Security frameworks such as SoFI and fault-injection attack emulation~\cite{b62,b63}, together with hardware fuzzing techniques such as RFUZZ, DirectFuzz, and HFL~\cite{b49,b103,b105}, show that targeted exploration can uncover non-trivial vulnerabilities, but they also highlight the need for more meaningful metrics that reflect which assets, threat classes, and microarchitectural states have actually been exercised. Existing vendor coverage infrastructures for emulation and current research prototypes provide useful ingredients, but practical, tool-friendly definitions of security coverage remain immature~\cite{report1,rad1}. Without such metrics, it is difficult to compare different emulation-based security flows, quantify the assurance gained from a given campaign, or argue that an SoC has been tested to a defensible security depth.

\section{Emerging Opportunities, Research Directions, and Open Problems}
\label{sec:research_directions}

Although current deployments of emulation-based security verification are still relatively ad hoc, underlying technology and ecosystem trends point to several promising directions through which emulation can become a more central pillar of hardware security assurance. This section highlights opportunities that build on the foundations of SoC security verification~\cite{b1,b2,b32}, fuzzing and penetration testing~\cite{b15,b16,b18,b19,b20,b42,b43,b44,b49,b50,b51}, and fault-injection-based vulnerability assessment~\cite{b23,b62,b63}. It also highlights research directions and open problems that emerge from the current state of the field, with particular emphasis on automation, adversarial campaign design, monitor generation, and evaluation methodology.

\begin{figure}[t]
    \centering
    \includegraphics[width=0.8\linewidth]{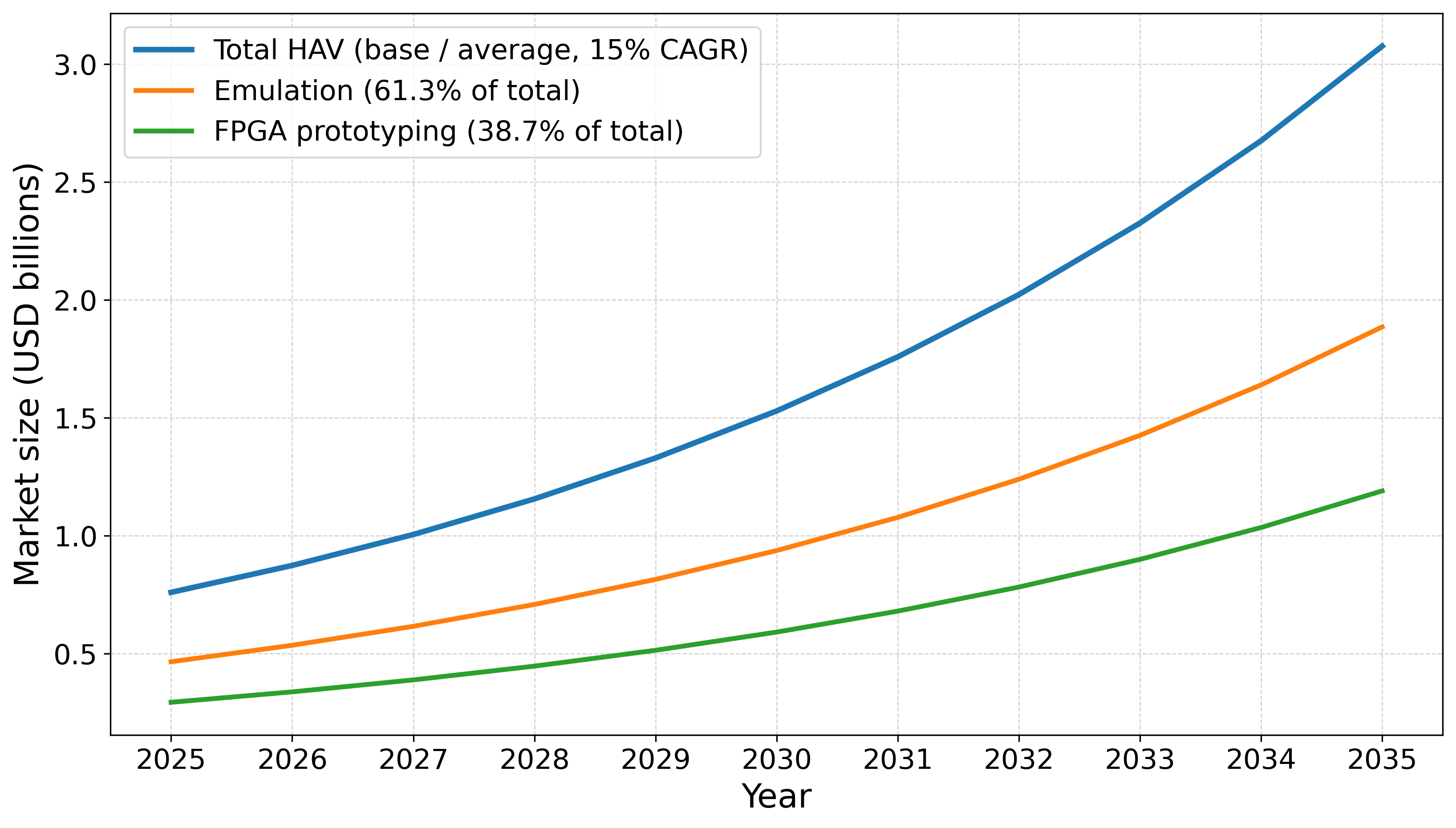}
    \caption{Forecasted growth of hardware-assisted verification (HAV) segments through 2035, highlighting increasing adoption of emulation- and prototyping-based verification.}
    \label{fig:projection}
\end{figure}

Figure~\ref{fig:projection} plots the projected growth of the global hardware-assisted verification (HAV) market using published market-sizing estimates and a constant platform split. The base case is anchored at USD~760.4~M in 2025 and projected to reach USD~3,076.4~M by 2035 (15.0\% CAGR), decomposed into hardware emulation (61.3\% share in 2025) and FPGA prototyping (38.7\%) using the same source~\cite{market1}. The overall magnitude and growth trend are consistent with an independent estimate reporting USD~641.5~M in 2024 and USD~1.5~B by 2030 (15.2\% CAGR)~\cite{market2}. For broader electronic design market context, we also cite the ESD Alliance (SEMI) EDMD release reporting total electronic system design industry revenue of USD~5.6~B in Q3~2025, including computer-aided engineering (CAE) revenue of USD~2,097.8~M~\cite{market3}.

\subsection{AI-Native Emulation and Autonomous Security Campaigns}

Recent work shows that machine learning and AI techniques can substantially improve SoC security verification, for example by guiding fuzzing, penetration testing, and test prioritization~\cite{b16,b44,b97}. Large language models (LLMs) extend this trend by reasoning over specifications, threat models, and security requirements, and can assist with vulnerability analysis, test generation, and artifact understanding~\cite{b114,b115,b116,b117}.

Emulation creates an opportunity to close the loop between AI-guided planning and high-fidelity execution. Instead of manually selecting assets and crafting campaigns, learning-based agents can propose security scenarios---for example privilege-escalation attempts, misconfiguration of debug features, malformed transaction sequences, or multi-stage attack paths---and map them to workloads that run realistic firmware and software stacks~\cite{b17,b18,b19,b16,b42,b103}. The resulting traces, monitor events, and security violations can then be used to refine the search strategy, prioritize suspicious behaviors, and generate stronger follow-on tests. A long-term direction is therefore not merely ``AI for security verification,'' but \emph{closed-loop emulation-driven security exploration} in which test generation, execution, and evidence analysis iteratively improve one another.

\subsection{Synergistic Fault Emulation and Hardware Fuzzing}

Emulation-based fault-injection frameworks~\cite{b62,b63} and hardware fuzzing techniques~\cite{b15,b17,b18,b16,b42,b100,b103} have so far evolved largely in parallel. Yet the two are naturally complementary. Fault campaigns provide precise control over \emph{where} and \emph{when} perturbations are introduced, while fuzzing explores large input and state spaces under rich behavioral feedback. A key research direction is to unify them within a single campaign framework.

For example, fuzzer feedback can be augmented with security-oriented observables such as information-flow violations~\cite{b12,b13,b32,b101,b102}, side-channel-sensitive timing behavior~\cite{b21,b22,b24,b25}, or assertion failures derived from threat models and security policies~\cite{b1,b27,b43,b44,b97}. Fault-injection engines can then focus on security-sensitive phases already identified by fuzzing, such as key handling, privilege transitions, or protection-domain crossings. Conversely, fault campaigns can reveal rare exception-handling paths or latent states that seed stronger fuzzing inputs. Because emulation supports both high-throughput execution and repeatable perturbation, it is a natural substrate for such coupled workflows.

\subsection{Scaling to Heterogeneous SoCs, 3D ICs, and Chiplet-Based Systems}

As SoCs evolve into heterogeneous platforms with accelerators, complex interconnects, and chiplet- or package-based integration, the attack surface increasingly spans multiple dies, IP vendors, and technology nodes~\cite{b26,b32,b33}. Prior work has shown that advanced integration introduces new avenues for both logical and physical attacks~\cite{b33}, while system-level studies emphasize the importance of validating shared resources and interconnect behavior~\cite{b26}.

Emulation can play an important role here, but its role must be stated carefully. Its natural strength is \emph{logical and protocol-level} modeling of multi-die systems, not detailed physical modeling of interposer parasitics or package effects. Accordingly, the most realistic research directions are to:
\begin{itemize}
    \item model chiplet- and SiP-based integration at the protocol and architectural level, including die-to-die interfaces and root-of-trust hardware modules~\cite{b84};
    \item study malformed transactions, protocol-level faults, and security policy violations at inter-die boundaries under realistic software workloads~\cite{b33,b62,b63};
    \item validate end-to-end properties such as isolation, secure boot, and key management across multi-die systems.
\end{itemize}
Detailed physical effects of interposers and package parasitics remain the domain of lower-level timing and physical analysis tools; the role of emulation is to expose the \emph{system-level security consequences} of die-to-die communication behavior rather than to replace physical modeling.

\subsection{Digital Twins and Continuous Security Assurance}

Digital-twin concepts are gaining traction as a means to model, monitor, and assess complex systems over long operational lifecycles~\cite{b29,b122}. From a hardware-security perspective, emulation can serve as the execution engine of a security-focused digital twin: a synchronized representation of the deployed system's hardware, firmware, and configuration that supports repeated security reassessment.

Such an emulation-backed twin could be used to reproduce field incidents, evaluate the impact of firmware updates or configuration changes, and test new attack techniques against legacy platforms that are still deployed but no longer under active design verification. Security frameworks such as SeVNoC and SoFI already illustrate how validation can be elevated from block-level checks to system-level behavior involving NoCs and fault-injection resilience~\cite{b26,b62,b63}. Extending these ideas toward persistent emulation-backed reassessment is especially relevant for \emph{automotive} platforms, where long product lifecycles and repeated software updates increase the need for ongoing security validation~\cite{b29}.

\subsection{Formal-Assisted Emulation Workflows}

Formal verification and emulation are strongest in different regimes. Formal methods are effective for proving localized properties such as access-control invariants, privilege transitions, and protocol safety, while emulation is effective for executing long software-driven scenarios at scale~\cite{b4,b9,b10,b11}. A practical and defensible research direction is therefore not a vague ``hybrid'' claim, but \emph{formal-assisted emulation} in which formal analysis strengthens specific parts of the emulation workflow.

One concrete example is \emph{property-guided monitor generation}. Suppose a formal property states that unprivileged software must never write a protected control register. That property can first be validated formally for a tractable subsystem, and then translated into a synthesizable checker that runs during emulation. The emulator can subsequently execute full firmware/OS workloads and flag any runtime sequence that violates the same property under realistic system context. In this workflow, formal methods provide rigor for the property itself, while emulation provides the scale and software realism needed to test it under realistic operation.

A second example is \emph{counterexample-guided campaign refinement}. Formal analysis can expose corner-case traces or hard-to-reach conditions for a small block; those traces can then seed longer emulation campaigns that explore whether similar conditions propagate into SoC-level failures once software, interrupts, DMA, and peripheral activity are included~\cite{b19,b37,b105}. This kind of formal-assisted emulation is concrete, useful, and aligned with existing assertion-based verification practice.

\subsection{Coverage, Benchmarks, and Standardization}

The lack of well-defined security coverage metrics and shared benchmarks is a recurring theme in SoC security verification~\cite{b2,b4,b11,b25,b97,b104,b113}. For emulation, this gap is even more pronounced: while functional coverage models such as line, toggle, FSM, and transaction coverage are well established~\cite{b11,b47}, they correlate only weakly with security goals such as asset exposure, privilege misuse, or side-channel observability. Existing emulation infrastructures and current security-verification frameworks provide useful ingredients for richer measurement, but practical, tool-friendly notions of security coverage for long-running emulation campaigns remain immature~\cite{b53,report1,rad1}.

\begin{figure*}[t!]
\centering
\includegraphics[width=1\textwidth]{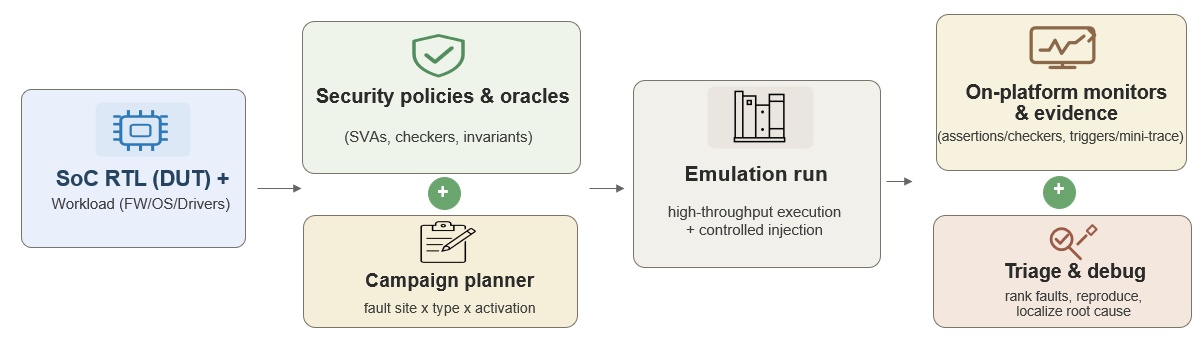}
\captionsetup{width=\textwidth}
\caption{Method template for emulation-based fault-injection vulnerability discovery:
property binding (SVAs/checkers), campaign planning (site/type/time), high-throughput emulation with controlled injection, and compact evidence for systematic triage and debug~\cite{b62,b63}.}
\label{fig:rp1_emfia_flow}
\end{figure*}

A clear research direction is to define metrics that capture both structural exploration (which blocks, states, and interfaces have been exercised under adversarial conditions) and semantic coverage (which assets, threat classes, and security requirements have been tested)~\cite{b1,b2,b25,b97,b104}. Coupled with standardized benchmark suites that package RTL designs, threat models, and representative attack scenarios~\cite{b15,b17,b18,b19,b16,b42,b43,b44,b97,b103}, such metrics would enable fairer comparison across tools and workflows and support more quantitative security sign-off.

\subsection{Research Problem 1: Emulation-Based Fault-Injection Vulnerability Discovery}
\label{sec:rp1}

Fault-injection attacks (FIAs) deliberately perturb hardware behavior to bypass checks, corrupt control flow, or induce information leakage, and have been demonstrated using voltage/clock glitching, electromagnetic interference, lasers, and focused ion beam (FIB) techniques~\cite{b23,b95,b96,b97,emfia}.
From a pre-silicon perspective, the key challenge is not whether FIAs matter, but how to assess resilience \emph{early}, \emph{systematically}, and under realistic workloads.

\subsubsection{Why emulation is a useful complement for early FIA assessment}

Pre-silicon FIA assessment is performed at multiple abstraction levels.
Physics- and layout-aware fault analysis remains important because it captures device-level and timing-level effects that cannot be represented faithfully at RTL~\cite{b98,b99,b100,b101}.
Such tools are essential for detailed analysis of critical nodes and for understanding how physical fault mechanisms map to circuit-level failures.
However, these analyses are computationally expensive and are typically applied later in the flow or to narrowed sets of scenarios.

This creates an opportunity for emulation-based assessment at the RTL level.
Emulation does not replace physics-based FIA tools; rather, it can be used earlier in the design flow to screen large numbers of signals, workloads, and activation windows under realistic HW/SW context.
The resulting set of suspicious or security-sensitive cases can then be prioritized for deeper gate-level, timing-aware, or layout-aware analysis.
In this sense, emulation complements simulation-based and physics-based FIA tools by adding campaign-scale exploration and realistic software-driven execution at an earlier design stage.
Recent work such as EmFIA further strengthens this view by showing that RTL-level emulation can support property-aware fault campaigns with dynamic signal forcing and online assertion checking, enabling much faster vulnerability assessment than exhaustive lower-level simulation alone~\cite{emfia}.

Three practical advantages make this complement valuable in modern SoCs:

\vspace{2pt}
\noindent\textbf{(1) Throughput for workload-driven screening.}
Meaningful security validation often requires long software executions (boot flows, drivers, OS services, interrupts, DMA, protocol stacks).
Emulation allows these workloads to be exercised over many more runs and perturbation points than would be feasible in pure RTL or gate-level simulation.

\vspace{2pt}
\noindent\textbf{(2) Fault-space reduction before expensive analysis.}
FIA resilience depends on \emph{where} a fault lands, \emph{what} it does, and \emph{when} it is activated.
Emulation can rapidly screen broad site/time combinations and identify a smaller set of critical cases for later, more detailed physical analysis.

\vspace{2pt}
\noindent\textbf{(3) Realistic system context.}
Many exploitable outcomes emerge only under realistic HW/SW interaction (privilege transitions, cache/TLB behavior, peripheral concurrency).
Emulation makes it possible to study these interactions early, before committing to more expensive low-level fault analyses.

\begin{table*}[t]
\centering
\caption{Positioning of representative pre-silicon FIA assessment approaches (qualitative).}
\label{tab:rp1_fia_methods}
\renewcommand{\arraystretch}{1.15}
\small
\begin{tabularx}{\textwidth}{@{}p{3.2cm} p{2.0cm} p{2.0cm} X p{2.4cm}@{}}
\toprule
\textbf{Approach} & \textbf{Model / level} & \textbf{Campaign scale} & \textbf{Typical strengths / constraints} & \textbf{Representative refs} \\
\midrule
Property-driven simulation-based FIA
& Gate / RTL sim.
& Low--Med
& Strong semantic visibility and property alignment, but slow for long HW/SW workloads and large (site$\times$time) sweeps.
& SoFI~\cite{b63} \\

Physically motivated fault analysis
& Timing/laser-oriented (sim.)
& Low
& Higher physical realism, but late-stage and expensive; difficult to iterate early and to scale campaigns under realistic software.
& HOST laser FIA~\cite{b101}; laser analysis~\cite{b98} \\

Campaign methodology / statistical FIA
& HDL robustness (sim.)
& Low--Med
& Useful campaign methodology, but still simulation-bound; throughput bottlenecks remain for SoC-scale software runs.
& EDCC~\cite{b100}; survey~\cite{b99} \\

FPGA-/acceleration-assisted testing
& FPGA / accelerated runs
& High
& High throughput, but often reduced RTL-level semantics/observability; debugging and property-centric attribution can be harder.
& RFUZZ~\cite{b49}; DirectFuzz~\cite{b103} \\

\textbf{RTL emulation campaigns (this work's framing)}
& RTL on emulator
& High
& Enables system-scale HW/SW execution with controlled injection and online property checking; supports practical pre-silicon vulnerability discovery loops.
& EmFIA framework~\cite{emfia}; SoFI context~\cite{b63} \\
\bottomrule
\end{tabularx}
\end{table*}

\subsubsection{What emulation enables for pre-silicon FIA assessment}

Emulation makes FIA campaigns practical when the evaluation requires (i) high-throughput execution, (ii) realistic HW/SW context (firmware, OS services, interrupts, DMA, device activity), and (iii) property-aware checking with controlled observability overhead~\cite{b94,b95}.
A particularly practical abstraction for early vulnerability discovery is a persistent, single-bit stuck-at fault injected at RTL, which captures controllable, permanent signal disruptions consistent with FIB-style edits while avoiding detailed analog/timing modeling complexity~\cite{b62,emfia}.
This supports a pre-silicon workflow where faults are injected at security-sensitive RTL signals and violations are detected online using embedded SystemVerilog Assertions (SVAs) and checkers that encode confidentiality, integrity, and availability expectations~\cite{b62,b63,emfia}.

\subsubsection{Workflow: property-aware fault-injection campaigns on an emulated RTL SoC}

Figure~\ref{fig:rp1_emfia_flow} presents a method-centric workflow template for emulation-driven FIA discovery:
(i) identify security-critical signals,
(ii) bind signals to security properties (SVAs/monitors),
(iii) plan a fault campaign (site/type/time),
(iv) execute faults with controlled runtime injection during emulation,
and (v) log violations and compact evidence for triage and debug~\cite{b62,b63,emfia}.
This organization emphasizes repeatable campaigns and systematic violation capture under realistic workloads, while preserving the option to hand off critical findings to lower-level FIA tools for more detailed analysis.

\subsubsection{Positioning relative to prior FIA assessment approaches}

Table~\ref{tab:rp1_fia_methods} positions representative pre-silicon FIA assessment styles.
The goal is not to restate results, but to clarify where RTL emulation campaigns sit in the design stack: earlier than layout-stage analyses, more property-aligned than bitstream-only injection, and more scalable than simulation-centric sweeps~\cite{b42,b43,b49,b50,emfia,b119,b84,b85}.

\begin{table*}[!t]
\centering
\caption{RL/ML-based fuzzing challenges and emulation-enabled opportunities.}
\label{tab:fuzzing_challenges}
\renewcommand{\arraystretch}{1.25}
\small
\begin{tabularx}{\textwidth}{@{}l X X l@{}}
\toprule
\textbf{Challenge domain} & \textbf{Limitation without emulation + RL/ML} & \textbf{Emulation-enabled opportunity} & \textbf{Representative refs} \\
\midrule
State-space exploration &
Naive mutation struggles against SoC complexity. Random or static strategies often fail to synthesize programs with complex control/data dependencies that reach deep states. &
Learning-based mutation/seed policies can focus exploration, while emulation provides the throughput required for many training iterations. &
\cite{b92,b105,b106,b107} \\

Feedback quality &
Pure software-side fuzzing feedback may miss security-relevant internal states or protocol-level conditions. &
Emulation can expose richer feedback sources such as assertions, monitors, counters, and selected internal security-state indicators. &
\cite{b42,b43,b92,b105} \\

Workload realism &
Interesting security behaviors often emerge only under realistic firmware, OS, driver, or peripheral interaction. &
Emulation enables these workloads to execute on RTL-speed hardware models while keeping the campaigns repeatable. &
\cite{b15,b16,b92,b105} \\

Training cost &
RL/ML requires many repeated executions to optimize policies and evaluate reward functions. &
High-throughput emulation reduces the execution bottleneck enough to make policy learning practical for hardware fuzzing. &
\cite{b92,b105,b106,b107} \\
\bottomrule
\end{tabularx}
\end{table*}

\subsection{Research Problem 2: Intelligent Hardware Fuzzing via Emulation and Reinforcement Learning}

Hardware fuzzing has emerged as a practical technique for exploring large input and state spaces in search of functional and security vulnerabilities. However, conventional coverage-guided fuzzing faces two persistent bottlenecks in hardware settings: limited execution throughput, which restricts long-running campaigns, and weak search efficiency, since naive mutation strategies often fail to reach deep SoC states or produce semantically meaningful stimuli~\cite{b15,b92,b105}. These limitations are especially severe for modern SoCs, where interesting behaviors frequently depend on long instruction sequences, protocol-compliant transactions, privilege transitions, interrupts, and complex HW/SW interactions.

Recent work suggests that reinforcement learning (RL) and related machine-learning (ML) techniques can improve fuzzing efficiency by learning mutation policies, seed prioritization strategies, and input-generation patterns that are better aligned with hard-to-reach states and vulnerability conditions~\cite{b105,b106,b107}. Approaches based on bandit methods, swarm optimization, sequence models, and deep RL indicate that learning-based guidance can outperform static or random mutation in both coverage and bug-finding effectiveness~\cite{b91,b105,b106,b107,b108}. Yet these gains come at a cost: such methods require large numbers of executions to generate feedback for training and policy refinement.

Table~\ref{tab:fuzzing_challenges} provides a structured summary of the main challenges in RL/ML-based hardware fuzzing and the corresponding ways in which emulation can mitigate execution, feedback, and scalability bottlenecks.

This creates a natural research problem for emulation-enabled security verification. Emulation provides the execution substrate needed to make RL/ML-guided fuzzing practical at the SoC level by supporting high-throughput, repeatable execution of RTL under realistic firmware and software context~\cite{b92,b105}. At the same time, RL/ML can make these campaigns more selective and more effective by directing exploration toward security-relevant states instead of relying on blind mutation alone. In this sense, emulation addresses the \emph{execution bottleneck}, while learning-based fuzzing addresses the \emph{search-efficiency bottleneck}.

A practical emulation-driven fuzzing loop would therefore combine three elements:
\begin{itemize}
    \item \textbf{High-throughput execution:} the emulator runs realistic RTL + firmware workloads at campaign scale;
    \item \textbf{Security-aware feedback:} coverage signals, assertions, taint/IFT monitors, differential checks, or software-visible outcomes provide reward or guidance signals;
    \item \textbf{Adaptive exploration:} RL/ML agents update mutation choices, seed selection, or attack-sequence generation based on observed outcomes.
\end{itemize}

This direction is particularly important for SoCs because many meaningful security states are software-mediated and cannot be exercised by shallow or purely random input streams~\cite{b42,b45,b102}. Effective fuzzing often requires structured stimuli such as valid instruction sequences, coordinated bus activity, MMIO accesses, DMA behavior, CSR interactions, and interrupt timing, together with practical oracles capable of detecting silent security violations such as privilege misuse or unintended information flow~\cite{b37,b42,b43,b45}. Emulation makes these longer, system-context-rich campaigns feasible, while learning-based policies make them increasingly targeted.

Accordingly, the core research challenge is not simply to speed up hardware fuzzing, but to build integrated emulation-based fuzzing frameworks in which throughput, feedback quality, and adaptive exploration reinforce one another. Such a framework could enable intelligent adversarial testing that is both scalable and security-aware, providing a path beyond static mutation and toward full-system vulnerability discovery at practical campaign scale~\cite{b92,b105,b106,b107,b109}.


\section{Concluding Remarks}

\label{sec:conclusion}

Security verification is increasingly constrained by scale, realism, and campaign volume: modern SoCs demand validation under long-running software stacks, complex concurrency, and adversarial stress.
Across these axes, hardware emulation offers a practical path to close the gap between what security techniques \emph{could} detect and what verification teams can \emph{afford} to run.

This paper positioned emulation-based security verification (ESV) as a design space, separated platform choices from workflow choices, and introduced a taxonomy that organizes ESV along security objectives, verification-engine integration, and instrumentation strategies.
We also highlighted key challenges---compile/mapping overheads, limited observability, trace scalability, and monitor automation---and identified research opportunities such as scalable emulation farms, security-aware coverage guidance, and more automated monitor synthesis.

Looking forward, the most impactful progress will come from treating ESV as a continuous, regression-driven workflow: security monitors and adversarial campaigns should run routinely alongside functional verification in pre-silicon infrastructure.
Doing so can make emulation not just an accelerator for late-stage debug, but a foundation for proactive and systematic SoC security validation.


\begin{thebibliography}{00}

\bibitem{b_SOC}
H.~Al-Shaikh, S.~Saha, S.~Kumar, F.~Farahmandi, and M.~Tehranipoor,
``Rethinking System-on-Chip Verification for Security Cross-Layer Interactions,''
\emph{IEEE Design \& Test}, 2025.

\bibitem{b1}
S.~Ray \emph{et al.},
``System-on-chip platform security assurance: Architecture and validation,''
\emph{Proc. IEEE}, vol.~106, no.~1, pp.~21--37, 2017.

\bibitem{b2}
X.~Chen \emph{et al.},
``Challenges and trends in modern SoC verification: A holistic perspective,''
\emph{IEEE Design \& Test}, vol.~34, no.~5, pp.~7--22, 2017.

\bibitem{b3}
R.~B.~Lee,
\emph{Security Basics for Computer Architects}.
Morgan \& Claypool, 2022.

\bibitem{b4}
S.~Witharana and S.~Ray,
``A survey on specification mining for hardware design and verification,''
\emph{ACM Comput. Surv.}, vol.~55, no.~9, pp.~1--38, 2022.

\bibitem{b5}
K.~Xiao and M.~Tehranipoor,
``Hardware Trojans: Lessons learned after one decade of research,''
\emph{ACM Trans. Design Autom. Electron. Syst.}, vol.~22, no.~1, pp.~1--23, 2016.

\bibitem{b6}
F.~Imeson \emph{et al.},
``A survey of hardware security research on SoCs, FPGAs, and discrete components,''
\emph{ACM J. Emerg. Technol. Comput. Syst.}, 2023.

\bibitem{b7}
C.~Xue \emph{et al.},
``Ten years of speculative execution attacks: A survey,''
\emph{ACM Comput. Surv.}, 2020.

\bibitem{b8}
B.~Aslan \emph{et al.},
``A comprehensive review of hardware Trojans and detection techniques,''
\emph{Electronics}, vol.~12, no.~6, p.~1333, 2023.

\bibitem{b21}
S.~Mangard, E.~Oswald, and T.~Popp,
\emph{Power Analysis Attacks: Revealing the Secrets of Smart Cards}.
Springer, 2007.

\bibitem{b22}
P.~Kocher, J.~Jaffe, and B.~Jun,
``Differential power analysis,''
in \emph{Advances in Cryptology---CRYPTO '99}, pp.~388--397, 1999.

\bibitem{b23}
E.~Biham and A.~Shamir,
``Differential fault analysis of secret key cryptosystems,''
in \emph{Advances in Cryptology---CRYPTO '97}, pp.~513--525, 1997.

\bibitem{b24}
D.~Agrawal, J.~R.~Rao, and P.~Rohatgi,
``The EM side-channel(s),''
in \emph{Cryptographic Hardware and Embedded Systems (CHES)}, pp.~29--45, 2002.

\bibitem{b9}
S.~Guo \emph{et al.},
``Scalable formal hardware verification using modularity and abstraction,''
\emph{IEEE Trans. Comput.-Aided Des. Integr. Circuits Syst.}, vol.~35, no.~7, pp.~1201--1214, 2016.

\bibitem{b10}
A.~Valmari,
``The state explosion problem,''
in \emph{Advanced Course on Petri Nets}, pp.~429--528, Springer, 1996.

\bibitem{b11}
H.~Chockler \emph{et al.},
``A practical approach to coverage in model checking,''
in \emph{Proc. CAV}, pp.~66--78, 2001.

\bibitem{b12}
W.~Hu \emph{et al.},
``Detecting and preventing information leakage in hardware designs,''
in \emph{Proc. ICCAD}, pp.~1--8, 2016.

\bibitem{b13}
W.~Hu \emph{et al.},
``Hardware information flow tracking: Overview and advances,''
\emph{IEEE Trans. Comput.-Aided Des. Integr. Circuits Syst.}, vol.~40, no.~4, pp.~1--20, 2021.

\bibitem{b14}
F.~Maragkou \emph{et al.},
``Information flow tracking for hardware security: A tutorial,''
\emph{ACM Trans. Design Autom. Electron. Syst.}, 2022.

\bibitem{b15}
T.~Trippel \emph{et al.},
``Fuzzing hardware like software,''
in \emph{Proc. USENIX Security}, pp.~3237--3254, 2022.

\bibitem{b16}
K.~Z.~Azar \emph{et al.},
``Fuzz, penetration, and AI testing for SoC security verification: Challenges and solutions,''
\emph{IEEE Design \& Test}, 2022.

\bibitem{b25}
K.~Gandolfi, C.~Mourtel, and F.~Olivier,
``Electromagnetic analysis: Concrete results,''
in \emph{Cryptographic Hardware and Embedded Systems (CHES)}, pp.~251--261, 2001.

\bibitem{b27}
W.~N.~N.~Hung and R.~Sun,
``Challenges in large FPGA-based logic emulation systems,''
in \emph{Proc. Int. Symp. Physical Design}, pp.~26--33, 2018.

\bibitem{b53}
A.~Koczor, {\L}.~Matoga, P.~Penkala, and A.~Pawlak,
``Verification approach based on emulation technology,''
in \emph{Proc. IEEE Int. Symp. Des. Diagn. Electron. Circuits Syst. (DDECS)}, 2016, pp.~1--6.

\bibitem{b26}
X.~Meng \emph{et al.},
``SeVNoC: Security validation of system-on-chip designs with NoC fabrics,''
\emph{IEEE Trans. Comput.-Aided Des. Integr. Circuits Syst.}, vol.~42, no.~2, pp.~672--682, 2023.

\bibitem{b33}
Y.~Xie, C.~Bao, C.~Serafy, T.~Lu, A.~Srivastava, and M.~Tehranipoor,
``Security and vulnerability implications of 3D ICs,''
\emph{IEEE Trans. Multi-Scale Comput. Syst.}, vol.~2, no.~2, pp.~108--122, 2016.

\bibitem{b17}
D.~Hur \emph{et al.},
``DIFFuzzRTL: Differential fuzz testing to find information leakage in RTL designs,''
in \emph{Proc. USENIX Security}, pp.~2969--2986, 2021.

\bibitem{b18}
M.~M.~Hossain \emph{et al.},
``SoCFuzzer: A fuzzing framework for SoC security verification,''
in \emph{Proc. DATE}, 2023.


\bibitem{b34}
S.~Bhunia, M.~S.~Hsiao, M.~Banga, and S.~Narasimhan,
``Hardware Trojan attacks: Threat analysis and countermeasures,''
\emph{Proc. IEEE}, vol.~102, no.~8, pp.~1229--1247, 2014.

\bibitem{b35}
M.~Xue, C.~Gu, W.~Liu, S.~Yu, and M.~O'Neill,
``Ten years of hardware Trojans: A survey from the attacker's perspective,''
\emph{IET Comput. Digit. Tech.}, vol.~14, no.~6, pp.~231--246, 2020.

\bibitem{b36}
\"O.~Aslan, S.~S.~Aktu{\u{g}}, M.~Ozkan-Okay, A.~A.~Yilmaz, and E.~Akin,
``A comprehensive review of cyber security vulnerabilities, threats, attacks, and solutions,''
\emph{Electronics}, vol.~12, no.~6, p.~1333, 2023.

\bibitem{b37}
W.~Hu, A.~Ardeshiricham, and R.~Kastner,
``Hardware information flow tracking,''
\emph{ACM Comput. Surv.}, vol.~54, no.~4, pp.~1--39, 2021.

\bibitem{b38}
S.~Maragkou and A.~Jantsch,
``Information flow tracking methods for protecting cyber-physical systems against hardware Trojans---a survey,''
\emph{arXiv:2301.02620}, 2022.

\bibitem{b44}
N.~F.~Dipu, M.~M.~Hossain, K.~Z.~Azar, F.~Farahmandi, and M.~Tehranipoor,
``FormalFuzzer: Formal verification assisted fuzz testing for SoC vulnerability detection,''
in \emph{Proc. Asia South Pac. Des. Autom. Conf. (ASP-DAC)}, 2024, pp.~355--361.

\bibitem{b42}
M.~M.~Hossain, A.~Vafaei, K.~Z.~Azar, F.~Rahman, F.~Farahmandi, and M.~Tehranipoor,
``SoCFuzzer: SoC vulnerability detection using cost function enabled fuzz testing,''
in \emph{Proc. Design, Autom. Test Eur. Conf. Exhib. (DATE)}, 2023, pp.~1--6.

\bibitem{b49}
S.~Laeufer, J.~Bachrach, A.~Kurth, and K.~Sen,
``RFUZZ: Coverage-directed fuzz testing of RTL on FPGAs,''
in \emph{Proc. IEEE/ACM Int. Conf. Comput.-Aided Des. (ICCAD)}, 2018, pp.~1--8.

\bibitem{b50}
H.~Al-Shaikh, A.~Vafaei, M.~M.~M.~Rahman, K.~Z.~Azar, F.~Rahman, F.~Farahmandi, and M.~Tehranipoor,
``Sharpen: SoC security verification by hardware penetration test,''
in \emph{Proc. Asia South Pac. Des. Autom. Conf. (ASP-DAC)}, 2023, pp.~579--584.

\bibitem{b85}
Cadence Design Systems,
``Palladium Emulation,''
[Online]. Available: \url{https://www.cadence.com/en_US/home/tools/system-design-and-verification/emulation-and-prototyping/palladium.html}.
Accessed: 2026-01-05.

\bibitem{b86}
Synopsys,
``ZeBu Emulation Systems,''
[Online]. Available: \url{https://www.synopsys.com/verification/emulation-prototyping/emulation.html}.
Accessed: 2026-01-05.

\bibitem{b87}
Siemens EDA,
``Veloce Strato CS emulation platform,''
[Online]. Available: \url{https://eda.sw.siemens.com/en-US/ic/hav/veloce-cs/strato-cs/}.
Accessed: 2026-01-05.

\bibitem{b110}
R.~Squiers,
``Co-modeling: A Powerful Capability for Hardware Emulation,''
Siemens Digital Industries Software, White Paper, 2019.
[Online]. Available:
\url{https://resources.sw.siemens.com/en-US/white-paper-co-modeling-a-powerful-capability-for-hardware-emulation/}

\bibitem{b111}
W.~Kim, H.~Park, H.~Kim, S.~B.~Choi, and S.~Kim,
``Early Software Development and Verification Methodology Using Hybrid Emulation Platform,''
in \emph{DVCon Proceedings}, 2017.
[Online]. Available:
\url{https://dvcon-proceedings.org/wp-content/uploads/early-software-development-and-verification-methodology-using-hybrid-emulation-platform.pdf}

\bibitem{b112}
R.~Yadav, M.~Khandelwal, S.~Kalbande, G.~Srivastava, and H.~Kim,
``Hybrid Emulation for faster Android Home screen bring up and Software Development,''
in \emph{DVCon Proceedings}, 2023.
[Online]. Available:
\url{https://dvcon-proceedings.org/wp-content/uploads/90995.pdf}

\bibitem{cad11}
Cadence Design Systems,
``Palladium emulation,''
Accessed: Mar. 18, 2026. [Online]. Available:
\url{https://www.cadence.com/en_US/home/tools/system-design-and-verification/emulation-and-prototyping/palladium.html}

\bibitem{hybrid}
Siemens EDA,
``Veloce apps,''
Accessed: Mar. 18, 2026. [Online]. Available:
\url{https://www.siemens.com/en-us/products/ic/hav/apps/}

\bibitem{cadence2}
Cadence Design Systems,
``Cadence Palladium Z1 enterprise emulation platform,''
Datasheet. Accessed: Mar. 18, 2026. [Online]. Available:
\url{https://www.cadence.com/content/dam/cadence-www/global/en_US/documents/tools/system-design-verification/palladium-z1-ds.pdf}

\bibitem{report1}
Synopsys,
``Synopsys extends leadership with enhanced verification continuum platform,''
May 30, 2019. Accessed: Mar. 18, 2026. [Online]. Available:
\url{https://news.synopsys.com/2019-05-30-Synopsys-Extends-Leadership-with-Enhanced-Verification-Continuum-Platform}

\bibitem{b62}
H.~Salmani and M.~Tehranipoor,
``Fault Injection Attack Emulation Framework for early evaluation of IC security,''
\emph{ACM Trans. Des. Autom. Electron. Syst.}, vol.~26, no.~6, 2021, Art.~no.~83.

\bibitem{b63}
Z.~Wang and M.~Tehranipoor,
``SoFI: Security property-driven vulnerability assessments of ICs against fault-injection attacks,''
\emph{IEEE Trans. Comput.-Aided Des. Integr. Circuits Syst.}, vol.~40, no.~11, pp.~2312--2325, 2021.

\bibitem{b67}
L.~Rizzatti and R.~Klein,
``Hardware Emulation for Software Validation (Part 2): Hybrid Emulation and Trace-Based Debug,''
\emph{Electronic Design}, May 5, 2017.
[Online]. Available:
\url{https://www.electronicdesign.com/technologies/test-measurement/article/21805014/hardware-emulation-for-software-validation-part-2-hybrid-emulation-and-trace-based-debug}.

\bibitem{b43}
M.~M.~Hossain, N.~F.~Dipu, K.~Z.~Azar, F.~Rahman, F.~Farahmandi, and M.~Tehranipoor,
``TaintFuzzer: SoC security verification using taint inference-enabled fuzzing,''
in \emph{Proc. IEEE/ACM Int. Conf. Comput.-Aided Des. (ICCAD)}, 2023, pp.~1--9.

\bibitem{b20}
N.~F.~Dipu \emph{et al.},
``FormalFuzzer: Formal-assisted fuzzing for SoC security,''
\emph{IEEE Trans. Comput.-Aided Des. Integr. Circuits Syst.}, 2024.

\bibitem{syn}
Siemens EDA,
``Veloce De-ICE App,''
Accessed: Mar. 18, 2026. [Online]. Available:
\url{https://www.siemens.com/en-us/products/ic/hav/apps/de-ice/}

\bibitem{rad1}
Cycuity,
``Security verification with Radix,''
Accessed: Mar. 18, 2026. [Online]. Available:
\url{https://cycuity.com/wp-content/uploads/2025/01/Cycuity_Security_Verification_2025.pdf}

\bibitem{b51}
H.~Al Shaikh, S.~Saha, K.~Z.~Azar, F.~Farahmandi, M.~Tehranipoor, and F.~Rahman,
``Re-Pen: Reinforcement learning-enforced penetration testing for SoC security verification,''
\emph{IEEE Trans. Very Large Scale Integr. (VLSI) Syst.}, 2024.

\bibitem{b61}
S.~Canakci, C.~Rajapaksha, L.~Delshadtehrani, A.~Nataraja, M.~B.~Taylor, M.~Egele, and A.~Joshi,
``ProcessorFuzz: Processor fuzzing with control and status registers guidance,''
in \emph{Proc. IEEE Int. Symp. Hardware Oriented Security Trust (HOST)}, 2023, pp.~1--12.

\bibitem{b32}
A.~Basak, S.~Bhunia, T.~Tkacik, and S.~Ray,
``Security assurance for system-on-chip designs with untrusted IPs,''
\emph{IEEE Trans. Inf. Forensics Security}, vol.~12, no.~7, pp.~1515--1528, 2017.

\bibitem{b103}
S.~Canakci, L.~Delshadtehrani, F.~Eris, M.~B.~Taylor, M.~Egele, and A.~Joshi,
``DirectFuzz: Automated Test Generation for RTL Designs using Directed Graybox Fuzzing,''
in \emph{Proc. ACM/IEEE Design Automation Conference (DAC)}, 2021, pp.~529--534,
doi: 10.1109/DAC18074.2021.9586289.

\bibitem{b105}
L.~Wu, M.~Rostami, H.~Li, and A.-R.~Sadeghi,
``HFL: Hardware Fuzzing Loop with Reinforcement Learning,''
in \emph{Proc. Design, Automation \& Test in Europe (DATE)}, 2025, pp.~1--7,
doi: 10.23919/DATE64628.2025.10993080.


\bibitem{b19}
M.~M.~Hossain \emph{et al.},
``TaintFuzzer: Taint-guided fuzzing for SoC security,''
in \emph{Proc. ICCAD}, 2023.

\bibitem{market1}
Future Market Insights,
``Hardware-Assisted Verification Market Size and Share Forecast Outlook 2025 to 2035,''
Aug.~1, 2025. Accessed: Mar.~11, 2026. [Online]. Available:
\url{https://www.futuremarketinsights.com/reports/hardware-assisted-verification-market}

\bibitem{market2}
Research and Markets,
``Hardware-Assisted Verification --- Global Strategic Business Report,''
2024. Accessed: Mar.~11, 2026. [Online]. Available:
\url{https://www.researchandmarkets.com/reports/6070921/hardware-assisted-verification-global}

\bibitem{market3}
ESD Alliance (a SEMI Technology Community),
``ESD Alliance Reports Electronic System Design Industry Posts \$5.6 Billion Dollars in Revenue in Q3 2025,''
Jan.~12, 2026. Accessed: Mar.~11, 2026. [Online]. Available:
\url{https://www.semi.org/en/semi-press-release/esd-alliance-reports-electronic-system-design-industry-posts-5.6-billion-dollars-in-revenue-in-q3-2025}


\bibitem{b97}
Z.~Chen, G.~Vasilakis, K.~Murdock, E.~Dean, D.~Oswald, and F.~D.~Garcia,
``VoltPillager: Hardware-based fault injection attacks against Intel SGX enclaves using the SVID voltage scaling interface,''
in \emph{Proc. USENIX Security Symposium}, 2021.

\bibitem{b100}
I.~Tuzov, D.~de Andr{\'e}s, and J.-C.~Ruiz,
``Accurate robustness assessment of HDL models through iterative statistical fault injection,''
in \emph{Proc. IEEE European Dependable Computing Conf. (EDCC)}, 2018, pp.~1--8.

\bibitem{b101}
N.~Pundir, H.~Li, L.~Lin, N.~Chang, F.~Farahmandi, and M.~Tehranipoor,
``Security properties driven pre-silicon laser fault injection assessment,''
in \emph{Proc. IEEE Int. Symp. Hardware Oriented Security and Trust (HOST)}, 2022, pp.~9--12.

\bibitem{b102}
J.~Xu, Y.~Liu, S.~He, H.~Lin, Y.~Zhou, and C.~Wang,
``MorFuzz: Fuzzing Processor via Runtime Instruction Morphing enhanced Synchronizable Co-simulation,''
in \emph{Proc. 32nd USENIX Security Symposium (USENIX Security)}, 2023, pp.~1307--1324.

\bibitem{b84}
M.~S.~U.~I.~Sami, T.~Zhang, A.~M.~Shuvo, M.~S.~U.~Haque, P.~E.~Calzada, K.~Z.~Azar, H.~M.~Kamali, F.~Rahman, F.~Farahmandi, and M.~Tehranipoor,
``Advancing trustworthiness in system-in-package: A novel root-of-trust hardware security module for heterogeneous integration,''
\emph{IEEE Access}, vol.~12, 2024, Art.~no.~3375874.

\bibitem{b29}
H.~Riggs, S.~Tufail, I.~Parvez, M.~Tariq, M.~A.~Khan, A.~Amir, K.~V.~Vuda, and A.~I.~Sarwat,
``Impact, vulnerabilities, and mitigation strategies for cyber-secure critical infrastructure,''
\emph{Sensors}, vol.~23, no.~8, p.~4060, 2023.

\bibitem{b104}
M.~Rostami, M.~Chilese, S.~Zeitouni, R.~Kande, J.~Rajendran, and A.-R.~Sadeghi,
``Beyond Random Inputs: A Novel ML-Based Hardware Fuzzing,''
\emph{arXiv preprint arXiv:2404.06856}, 2024.

\bibitem{b47}
F.~Solt, K.~Ceesay-Seitz, and K.~Razavi,
``Cascade: CPU fuzzing via intricate program generation,''
in \emph{Proc. USENIX Security Symp.}, 2024, pp.~5341--5358.

\bibitem{b95}
K.~Murdock, D.~Oswald, F.~D.~Garcia, J.~V.~Bulck, D.~Gruss, and F.~Piessens,
``Plundervolt: Software-based fault injection attacks against Intel SGX,''
in \emph{Proc. IEEE Symp. Security and Privacy (S\&P)}, 2020.

\bibitem{b96}
N.~Timmers and C.~Mune,
``Escalating privileges in Linux using voltage fault injection,''
in \emph{Proc. Workshop Fault Diagnosis and Tolerance in Cryptography (FDTC)}, 2017.

\bibitem{emfia}
T.~Rahman, S.~Saha, S.~K.~Saha, F.~Farahmandi, and M.~Tehranipoor,
``EmFIA: A novel emulation-based fault injection vulnerability assessment framework at RTL level,''
in \emph{Proc. IFIP/IEEE 33rd Int. Conf. Very Large Scale Integration (VLSI-SoC)},
2025, pp.~1--5, doi: 10.1109/VLSI-SoC64688.2025.11421747.

\bibitem{b98}
A.~Vasselle, A.~Morisset, M.~Flottes, and B.~Rouzeyre,
``Laser-induced fault injection: A practical analysis,''
in \emph{Proc. IEEE Int. On-Line Testing Symp. (IOLTS)}, 2018.

\bibitem{b99}
M.~Eslami, B.~Ghavami, M.~Raji, and A.~Mahani,
``A survey on fault injection methods of digital integrated circuits,''
\emph{Integration}, vol.~71, pp.~154--163, 2020.

\bibitem{b94}
J.~Li, B.~Zhao, and C.~Zhang,
``Fuzzing: a survey,''
\emph{Cybersecurity}, vol.~1, Art.~no.~6, Jun.~2018, doi: 10.1186/s42400-018-0002-y.

\bibitem{b92}
S.~Xiang \emph{et al.},
``Verification of Hardware and Software with Fuzzing: Challenges and Opportunities,''
in \emph{Proc. ACM/IEEE}, 2022, doi: 10.1145/3563768.3565549.

\bibitem{b106}
V.~Gohil, R.~Kande, C.~Chen, A.-R.~Sadeghi, and J.~Rajendran,
``MABFuzz: Multi-Armed Bandit Algorithms for Fuzzing Processors,''
\emph{arXiv preprint arXiv:2311.14594}, 2023.

\bibitem{b107}
C.~Chen, V.~Gohil, R.~Kande, A.-R.~Sadeghi, and J.~Rajendran,
``PSOFuzz: Fuzzing Processors with Particle Swarm Optimization,''
\emph{arXiv preprint arXiv:2307.14480}, 2023.

\bibitem{b91}
L.~Wu \emph{et al.},
``GenHuzz: An Efficient Generative Hardware Fuzzer,''
in \emph{Proc. USENIX Security Symposium}, 2025.
[Online]. Available: \url{https://www.usenix.org/system/files/usenixsecurity25-wu-lichao.pdf}.

\bibitem{b108}
R.~G\"otz, C.~Sendner, N.~Ruck, M.~Rostami, A.~Dmitrienko, and A.-R.~Sadeghi,
``RLFuzz: Accelerating Hardware Fuzzing with Deep Reinforcement Learning,''
in \emph{Proc. IEEE Int. Symp. Hardware Oriented Security and Trust (HOST)}, 2025, pp.~358--369,
doi: 10.1109/HOST64725.2025.11050051.

\bibitem{b45}
J.~Hur, S.~Song, D.~Kwon, E.~Baek, J.~Kim, and B.~Lee,
``DifuzzRTL: Differential fuzz testing to find CPU bugs,''
in \emph{Proc. IEEE Symp. Security Privacy (SP)}, 2021, pp.~1286--1303.

\bibitem{b109}
N.~N.~Mondol, A.~Vafaei, K.~Z.~Azar, F.~Farahmandi, and M.~Tehranipoor,
``RL-TPG: Automated Pre-Silicon Security Verification through Reinforcement Learning-Based Test Pattern Generation,''
in \emph{Proc. Design, Automation \& Test in Europe Conf. \& Exhibition (DATE)}, 2024, pp.~1--6,
doi: 10.23919/DATE58400.2024.10546566.


\bibitem{b113}
R.~Kibria, F.~Farahmandi, and M.~Tehranipoor,
``A Survey on SoC Security Verification Methods at the Pre-silicon Stage,''
\emph{IACR Cryptology ePrint Archive}, 2024.

\bibitem{b114}
P.~E.~Calzada, Z.~Ibnat, T.~Rahman, K.~Kandula, D.~Lu, S.~K.~Saha, F.~Farahmandi, and M.~Tehranipoor,
``VerilogDB: The Largest, Highest-Quality Dataset with a Preprocessing Framework for LLM-based RTL Generation,''
\emph{arXiv preprint arXiv:2507.13369}, 2025.

\bibitem{b115}
A.~Ayalasomayajula, N.~Farzana, M.~Tehranipoor, and F.~Farahmandi,
``Automatic Asset Identification for Assertion-based SoC Security Verification,''
\emph{IEEE Trans. Comput.-Aided Des. Integr. Circuits Syst.}, 2024.

\bibitem{b116}
N.~Farzana, A.~Ayalasomayajula, M.~Tehranipoor, and F.~Farahmandi,
``AGILE: Automated Assertion Generation to Detect Information Leakage,''
\emph{IEEE Trans. Inf. Forensics Security}, 2024.

\bibitem{b117}
S.~Rajendran, N.~Farzana, S.~Tarek, H.~Kamali, F.~Farahmandi, and M.~Tehranipoor,
``Exploring the Abyss? Unveiling Systems-on-Chip Hardware Vulnerabilities beneath Software,''
\emph{IEEE Trans. Inf. Forensics Security}, 2024.

\bibitem{b118}
M.~M.~Hossain, K.~Z.~Azar, F.~Rahman, F.~Farahmandi, and M.~Tehranipoor,
``Fuzzing for Automated SoC Security Verification: Challenges and Solutions,''
\emph{IEEE Design \& Test}, 2024.

\bibitem{b119}
K.~Z.~Azar, M.~M.~Hossain, A.~Vafaei, H.~Al~Shaikh, N.~Mondol, F.~Rahman, M.~Tehranipoor, and F.~Farahmandi,
``Fuzz, Penetration, and AI Testing for SoC Security Verification: Challenges and Solutions,''
\emph{IACR Cryptology ePrint Archive}, 2022.

\bibitem{b120}
N.~Pundir, J.~Park, F.~Farahmandi, and M.~Tehranipoor,
``Power Side-Channel Leakage Assessment Framework at Register-Transfer Level,''
\emph{IEEE Trans. Very Large Scale Integr. (VLSI) Syst.}, 2022.

\bibitem{b121}
A.~Mazumder, T.~Zhang, F.~Farahmandi, and M.~Tehranipoor,
``A Comprehensive Survey on Non-Invasive Fault Injection Attacks,''
\emph{IACR Cryptology ePrint Archive}, 2023.

\bibitem{b122}
H.~A.~Shaikh, M.~B.~Monjil, S.~Chen, F.~Farahmandi, N.~Asadi, M.~Tehranipoor, and F.~Rahman,
``Digital Twin for Secure Semiconductor Lifecycle Management: Prospects and Applications,''
\emph{IACR Cryptology ePrint Archive}, 2022.




\end{thebibliography}
\end{document}